\documentclass[10pt,preprint]{aastex}

\shorttitle{Rotochemical Heating in MSPs}
\shortauthors{Fern\'andez \& Reisenegger}

\begin{document}
\title{Rotochemical Heating in Millisecond Pulsars. \\Formalism and
Non-superfluid case.}
\author{Rodrigo Fern\'andez and  Andreas Reisenegger\altaffilmark{1}}
\affil{Departamento de Astronom\'\i a y Astrof\'\i sica, Pontificia Universidad
Cat\'olica de Chile,\\
Casilla 306, Santiago 22, Chile.}
\altaffiltext{1}{E-mail: areisene@astro.puc.cl}

\begin{abstract}
Rotochemical heating originates in a departure from beta
equilibrium
due to spin-down compression in a rotating neutron star. The main
consequence is that the star eventually
arrives at a quasi-equilibrium state, in which the thermal photon
luminosity depends only on the current value of the spin-down
power, which is directly measurable.
Only in millisecond pulsars the spin-down power remains high long
enough for this state to be reached with a substantial luminosity.
We report an extensive study of the effect of this heating
mechanism on the thermal evolution of millisecond pulsars,
developing a general formalism in the slow-rotation approximation
of general relativity that takes the spatial structure of the star
fully into account, and using a sample of realistic equations of
state to solve the non-superfluid case numerically. We show that
nearly all observed millisecond pulsars are very likely to be in
the quasi-equilibrium state. Our predicted quasi-equilibrium
temperatures for PSR J0437-4715 are only 20\% lower than inferred
from observations. Accounting for superfluidity should increase
the predicted value.
\end{abstract}


%%%%%%%%%%%%%%%%%%%%%%%%%%%%%%%%%%%%%%%%%%%%%%%%%%%%%%%%%%%%%%%
\section{INTRODUCTION}

Neutron star cooling is an important tool for the study of dense
matter. By comparing cooling models with observations of thermal
emission from these objects, we can gain insight into the equation
of state (EOS) of dense matter, the signatures of exotic
particles, superfluid energy gaps, and magnetic field properties
[for a recent review, see \citet{yak04}]. A neutron star loses the
thermal energy with which it was born initially through neutrino
emission, and later through photon emission, the change between
these two stages occurring about $10^5$ yr after birth.

Several heating mechanisms may become important during late stages
of the evolution of a neutron star, most of them related to the
delayed adjustment to the progressively changing equilibrium state
as the rotation slows down. These include the dissipation of
rotational energy due to interactions between superfluid and
normal components of the star \citep{shibazaki89} and release of
strain energy stored by the solid crust due to spin-down
deformation \citep{cheng92}.

Another of these mechanisms is {\it rotochemical heating}
\citep{reisenegger95}, which has its origin in deviations from
beta equilibrium due to spin-down compression. As a neutron star
spins down, the centrifugal force acting on each fluid element
diminishes, changing the local value of the pressure. Since the
composition of neutron star matter in beta equilibrium is a
one-parameter function, this compression results in a displacement
of the equilibrium concentration of each particle species.
Reactions which change the chemical composition are in charge of
driving the system to the new equilibrium configuration. But if
the rate at which reactions do this task is slower than the change
of the equilibrium concentrations due to spin-down compression,
the system is permanently out of chemical equilibrium. This
implies an excess of energy, which is dissipated by enhanced
neutrino emission and heat generation.

After its introduction, this heating mechanism was studied by
several authors: \citet{cheng96} applied it to quark stars,
\citet{reisenegger97} made an order-of-magnitude estimate of the
effects of superfluidity, and \citet{iidasato97} studied the
heating due to compositional transitions in the crust due to
spin-down compression. All studies agree that rotochemical heating
is particularly important for old neutron stars with fast rotation
and low magnetic fields, features that are characteristic of
millisecond pulsars (MSPs).

The most striking prediction associated with rotochemical heating
is that, if the spin-down timescale is substantially longer than
any other timescale involved (with the likely exception of
magnetic field decay), the star arrives at a
\emph{quasi-equilibrium state}, in which the temperature depends
only on the current, slowly changing value of $\Omega
\dot{\Omega}$
(the product of the angular velocity and its time derivative),
proportional to the spin-down power, and not on the star's
previous history \citep{reisenegger95}. This provides a simple way to
constrain the physics involved in theoretical models, once the
spin parameters and observed surface temperature of a MSP are
known.

Although the qualitative behavior of the thermal evolution of
neutron stars with rotochemical heating is known, all previous
studies made only order-of-magnitude estimates, ignoring the
spatial structure of the star and thus not offering reliable
predictions to be compared with observations. Also unknown is the
dependence of the thermal evolution on EOS and stellar mass.

In this work, we calculate the thermal evolution of MSPs with
rotochemical heating, taking the structure of the star fully into
account in the frame of general relativity, and using realistic
EOSs of dense matter. In order to do so, we develop a general
formalism to treat the evolution of the temperature and departures
of chemical equilibrium. As a first approach, we have chosen the
simplest possible core composition, namely neutrons, protons,
electrons, and muons ($npe\mu$ matter), to see if we can explain
observations with it, before invoking exotic particles. The main
difficulty of treating the spatial structure of the star is the
lack of an expression for the spin-down compression in the frame
of general relativity. We develop such an expression with the aid
of the slow-rotation approximation of \citet{hartle67}. The
astrophysical situation to be modelled is the stage after the
accretion-driven spin-up of the pulsar has finished.

Although it is generally believed that crusts and cores of neutron
stars have superfluid components (e.g., \citealt{yak04}), in this
work we ignore superfluidity. Our plan is to make an extensive
study of rotochemical heating in non-superfluid stars and compare
results with observations, in order to assess whether invoking
superfluidity (whose parameters are currently very uncertain) is
necessary to explain observations.

The structure of this paper is the following: In \S2, we develop
our general formalism, as well as the expression for the spin-down
compression rate. In \S3, we detail the input necessary for
numerical calculations (EOS, reactions, heat capacities, etc.). In
\S4, we describe our results, comparing our predictions with the
recent detection of likely thermal ultraviolet emission from the
nearest MSP, PSR J0437-4715
\citep*{kargaltsev04}, and the lack of optical emission from the
even closer ``classical'' pulsar PSR J0108-1413
\citep*{mignani03}. We also list some other MSPs for which the
rotochemical heating luminosity might become detectable.
A short summary of our main conclusions is given in \S5.

%%%%%%%%%%%%%%%%%%%%%%%%%%%%%%%%%%%%%%%%%%%%%%%%%%%%%%%%%%%%%%%%%%%%%%%%%%%%
\section{THEORETICAL FRAMEWORK}
\label{sec:framework}

%---------------------------------------------------------------------------
\subsection{Basic Equations}

It is conventional in neutron star cooling calculations to divide
the star into two regions: a nearly isothermal \emph{interior},
which ranges from the center out to density $\rho_b \sim 10^{10}$
g cm$^{-3}$, and a thin \emph{envelope}, which ranges from
$\rho_b$ to the surface and where a strong temperature gradient
exists \citep*{gpe83}. Since we are modelling the thermal
evolution of a MSP long after accretion has stopped, it is safe to
assume that thermal relaxation from an initial non-uniform
internal temperature profile has already occurred, so that the
redshifted internal temperature,
\begin{equation}
\label{eq:T_int}
T_\infty = T(r)e^{\Phi(r)},
\end{equation}
is uniform (see, e.g., \citealt{glend97}). Here, $g_{tt} =
-e^{2\Phi}$ is the time component of the metric of a non-rotating
reference star, of which $r$ is the radial spherical coordinate.
Although spherical symmetry is broken for a rotating star, we
describe the latter as a perturbation to the corresponding
non-rotating star (with the same total baryon number). We
establish a Lagrangian correspondence between the surfaces of
constant $r$ in the non-rotating star with the constant-pressure
surfaces of its rotating counterpart, on which all local
thermodynamic quantities will be shown to be constant (see
\S~\ref{sec:perturbation}).
The evolution of the internal temperature is given by the thermal
balance equation \citep{thorne77}, which for an isothermal
interior is given by
\begin{equation}
\label{eq:dot_T}
\dot{T}_\infty = \frac{1}{C}\left[ L^{\infty}_H - L^{\infty}_\nu -
L^{\infty}_\gamma \right],
\end{equation}
where $C$ is the total heat capacity of the star, $L^{\infty}_H$ is the total
power released by heating
mechanisms,
$L^{\infty}_\nu$ is the total neutrino luminosity, and $L^{\infty}_\gamma$ is
the photon luminosity.
These quantities are calculated as
\begin{eqnarray}
\label{eq:capcal}
C & = & \sum_i\int dV c_{V,i}, \\
\label{eq:L_H}
L_H^\infty & = & \int dV Q_H e^{2\Phi},\\
\label{eq:L_nu}
L_\nu^\infty & = & \int dV Q_\nu e^{2\Phi},\textrm{ and}\\
\label{eq:L_gamma}
L_\gamma^\infty & = & 4\pi \sigma R^2 T_s^4 e^{2\Phi_s} = 4\pi \sigma
R_\infty^2 (T_s^\infty)^4,
\end{eqnarray}
respectively, where $dV=4\pi r^2 \sqrt{g_{rr}}dr$ is the proper
volume element, $c_{V,i}$ is the specific heat (heat capacity per
unit volume) of each particle species, $Q_\nu$ is the total
neutrino emissivity contributed by reactions, $Q_H$ is the total
heating rate per unit volume, $\sigma$ is the Stefan-Boltzmann
constant, $R$ is the stellar coordinate radius, $\Phi_s =
\Phi(R)$, $R_\infty = Re^{-\Phi_s}$ is the effective radius as
measured from infinity, and $T_s^\infty$ the redshifted effective
temperature. The surface temperature $T_s$ is obtained from the
internal temperature by assuming an envelope model
\citep{gpe83,potetal97}.

%----------------------------------------------------
Since the neutrino emissivity and heating rate are modified when
the neutron star is out of chemical equilibrium
\citep{haensel92,reisenegger95}, the evolution of the temperature
depends on how strongly it departs from the beta equilibrium
state. For $npe\mu$ matter, this departure can be quantified by
the chemical imbalances \citep{haensel92}:
\begin{eqnarray}
\label{eq:eta1}
\eta_{npe} & = & \delta \mu_n - \delta \mu_p - \delta \mu_e,\\
\label{eq:eta2}
\eta_{np\mu} & = & \delta \mu_n - \delta \mu_p - \delta \mu_\mu,
\end{eqnarray}
where $\delta \mu_i = \mu_i - \mu_i^{eq}$ is the deviation from
the equilibrium chemical potential of species $i$, at a given
pressure. Since diffusion timescales are short compared with the
evolutionary timescales to be considered \citep{reisenegger97}, we
assume uniform redshifted chemical potential deviations throughout
the core:
\begin{equation}
\label{eq:delta_mu_infty}
\delta \mu_i^\infty \equiv \delta \mu_i(r) e^{\Phi(r)}.
\end{equation}
To obtain the time evolution of the chemical imbalances, we start by writing
down the chemical potential
of each particle species as a function of the number density of all particles:
$\mu_i = \mu_i(\left\{ n_j \right\})$.
We assume small departures from chemical equilibrium, which can be
expressed by requiring that $|\delta \mu_i| \ll \mu_i^{eq}$. In
this approximation, the departures from the equilibrium particle
number densities $\delta n_i = n_i - n_i^{eq}$ are related to the
$\delta \mu_i$ by
\begin{equation}
\label{eq:dni_dmui} \delta n_i = \sum_j \frac{\partial
n_i}{\partial \mu_j} \delta \mu_j,
\end{equation}
where the partial derivatives are evaluated at the beta
equilibrium state. To eliminate the effect of particle diffusion
between different regions of the star, we integrate
equation~(\ref{eq:dni_dmui}) over regions where free particles
exist. After integration, we obtain the deviation from the
equilibrium number of particles $\delta N_i$ as a function of the
redshifted chemical potential lags:
\begin{equation}
\label{eq:dNi_dmui}
\delta N_i =  \sum_j B_{ij} \delta \mu_j^\infty,
\end{equation}
with
\begin{equation}
\label{eq:Bij_def}
B_{ij}  =  \int_{core} dV \frac{\partial n_i}{\partial \mu_j} e^{-\Phi},
\end{equation}
where we have used equation~(\ref{eq:delta_mu_infty}). Since the
$B_{ij}$ do not depend on time, we invert and take the time
derivative of equation~(\ref{eq:dNi_dmui}), obtaining the
evolution of the $\delta \mu^\infty_i$:
\begin{equation}
\delta \dot{\mu}_i^\infty = \sum_j \left( B^{-1} \right)_{ij} \delta \dot{N}_j.
\end{equation}
The rate of change of $\delta N_i$ is given by
\begin{equation}
\label{eq:dNi_def}
\delta \dot{N}_i = \dot{N}_i - \dot{N}_i^{eq},
\end{equation}
where
\begin{equation}
\label{eq:dot_Ni}
\dot{N}_i = \int_{core} dV e^{\Phi}\sum_\alpha \Delta \Gamma^i_\alpha
\end{equation}
is the change in the total number of particles of species $i$ due
to reactions \citep{thorne77}. Here, $\Delta \Gamma^i_\alpha$ is
the net creation rate of particles of species $i$ per unit volume
due to reaction $\alpha$. It can be checked that the $\dot{N_i}$
satisfy baryon number and charge conservation. The quantity
$\dot{N}_i^{eq}$ is the change in the equilibrium value of the
total number of particles of species $i$, which depends on the
spin-down compression rate. We defer its calculation to the next
subsection. The evolution of the redshifted chemical imbalances
$\eta^\infty= \eta(r)e^{\Phi}$ follows from
equations~(\ref{eq:eta1}), (\ref{eq:eta2}), and
(\ref{eq:delta_mu_infty}):
\begin{eqnarray}
\label{eq:dot_eta1}
\dot{\eta}^\infty_{npe} = \delta \dot{\mu}^\infty_n - \delta \dot{\mu}^\infty_p
- \delta \dot{\mu}^\infty_e\\
\label{eq:dot_eta2}
\dot{\eta}^\infty_{np\mu} = \delta \dot{\mu}^\infty_n - \delta
\dot{\mu}^\infty_p - \delta \dot{\mu}^\infty_\mu
\end{eqnarray}
Equations~(\ref{eq:dot_T}), (\ref{eq:dot_eta1}), and
(\ref{eq:dot_eta2}) give a complete description of the thermal
evolution of a neutron star with rotochemical heating and $npe\mu$
composition, given an expression for $\dot{N}_i^{eq}$.

%---------------------------------------------------------------------------
\subsection{Lagrangian Spin-down Compression Rate}
\label{sec:perturbation}

For slow enough rotation frequencies, the deviation of the star
from the non-rotating configuration can be treated as a small
perturbation, which we describe in terms of a Lagrangian
formalism. Using the fact that the total baryon number $A$ of a
star is conserved, we can describe its interior in terms of
surfaces of constant pressure $P$ enclosing a fixed number of
baryons $N$ (e.g., the surface $P=0$ encloses $A$ baryons inside
it). As the stellar rotation rate changes, these surfaces will
readjust their shapes, with a corresponding change in $P$, but
keeping $N$ constant. Since there is a one-to-one relation between
the enclosed number of baryons $N$ and the radial coordinate of
the non-rotating configuration $r$ (for fixed $A$), we use them
interchangeably to identify a given surface, relating them by $dN
= 4\pi r^2 \sqrt{g_{rr}}ndr$.

The underlying assumption is that, on a surface of constant
pressure, the other thermodynamical quantities are also constant.
This is straightforward for neutron-star matter in beta
equilibrium (described by a barotropic equation of state), but not
so obvious for departures from this state (which are not
necessarily barotropic, as the pressure-density relation will also
depend on the local particle abundances). We show in
Appendix~\ref{sec:gradients} that, in a uniformly rotating,
perfect-fluid star in hydrostatic equilibrium, surfaces of
constant pressure coincide with those of constant energy density
and gravitational redshift, regardless of whether the equation of
state is barotropic.\footnote{The validity of the assumption of a
perfect fluid is also discussed in Appendix~\ref{sec:gradients}. }
This, together with diffusive equilibrium (uniform redshifted
chemical potentials everywhere in the stellar interior), ensures
that the chemical composition on each surface is also constant.

To quantify the spin-down compression rate and thus
$\dot{N}^{eq}_i$, we need an expression for the change in pressure
with rotation frequency on each surface enclosing a constant $N$,
keeping $A$ constant. \citet{hartle67} developed a perturbative
approach in the frame of general relativity, relating quantities
on rotating and non-rotating stars with the same central energy
density $\rho_c$ (but, therefore, different $A$), by expanding the
metric of an axially symmetric non-rotating star in even powers of
the angular velocity $\Omega$. This perturbative approach is valid
for stars rotating at frequencies much smaller than the Kepler
frequency $\Omega_K$, at which mass shedding from the equator
occurs. For realistic equations of state, this limiting frequency
can be well approximated by means of empirical formulae
\citep*{lasota96}. The corresponding Kepler periods, $P_K$, are
lower than the shortest MSP periods measured to date
[see, e.g., The ATNF Pulsar
Catalogue\footnote{http://www.atnf.csiro.au/research/pulsar/psrcat/expert.html}
\citep{atnf}]. Our task is to express our spin-down compression
rate at constant $A$ in terms of Hartle's results, which assume
constant $\rho_c$.

%----------------------------------------------------------
If the baryon number enclosed by a surface of constant pressure $P$ in a
rotating star is
$N = N(A, P, \Omega^2)$, some partial-derivative manipulation yields
\begin{eqnarray}
\label{eq:dPdOmega2_final}
\left(\frac{\partial P}{\partial \Omega^2}\right)_{N, A} & = & -\frac{(\partial
N/\partial \Omega^2)_{P, A}}{(\partial N/ \partial P)_{\Omega^2, A}}\nonumber
\\
& = & -\frac{1}{ \left( {\partial N}/{\partial P} \right)_{\Omega^2, A} }
\left\{  \left( \frac{\partial N}{\partial \Omega^2} \right)_{P,\rho_c} -
\left( \frac{\partial N}{\partial \rho_c} \right)_{P, \Omega^2}\frac{\left(
{\partial A}/{\partial \Omega^2} \right)_{\rho_c}}{\left( {\partial
A}/{\partial \rho_c} \right)_{\Omega^2}}  \right\},
\end{eqnarray}
where the subscripts denote quantities which are kept constant in
carrying out the differentiation. We note that $(\partial
A/\partial \Omega^2)_{\rho_c}$ and $(\partial A/ \partial
\rho_c)_{\Omega^2}$ are just $(\partial N/\partial \Omega^2)_{P,
\rho_c}$ and $(\partial N/\partial \rho_c)_{P, \Omega^2}$,
respectively, evaluated at the surface of the star ($P=0$).
Also, since we are perturbing around the non-rotating
configuration, we can evaluate all the derivatives at $\Omega=0$.

%---------------------------------------------------------------
\citet{hartle67} found that the lowest-order change in a given
stellar property with rotation is linear in the squared angular
frequency $\Omega^2$, for a star with fixed central density. We
can thus write
\begin{equation}
\label{eq:dNdOmega2_n0}
\left( \frac{\partial N}{\partial \Omega^2}\right)_{P, \rho_c} =
\frac{N(\rho_c, P, \Omega^2) - N(\rho_c, P, 0)}{\Omega^2},
\end{equation}
and consider it independent of $\Omega^2$.
To find the number of baryons enclosed by a surface of constant pressure, we
recall the baryon number
conservation law \citep{misnersharp64}, and write
\begin{equation}
\label{eq:baryon_law}
N = \int_V (-g)^{1/2} u^t n d^3x,
\end{equation}
where $g$ is the determinant of the metric, $u^t$ is the time
component of the fluid's four-velocity, $n$ is the rest frame
baryon number density, and $V$ is the spatial volume enclosed by
the constant pressure surface (throughout this section, we use
$G=c=1$). For the non-rotating case, we have
\begin{equation}
   \label{eq:N_norot}
N(\rho_c, P, 0) = \int_0^{r(\rho_c, P)}4\pi y^2 e^{\Lambda(\rho_c, y)}
n(\rho_c, y) dy,
\end{equation}
where $g_{rr} = e^{2\Lambda} = (1 - 2M/r)^{-1}$ is the radial component of the
non-rotating metric.
For the rotating case, one has to express the perturbed metric of
\citet{hartle67} in the non-rotating
coordinate system. Keeping terms to order $\Omega^2$, we get
\begin{eqnarray}
\label{eq:g_rot}
\left( -g \right)^{-1/2} & = & e^{\Phi + \Lambda}r^2\sin{\theta}\left[ 1 +
\frac{m}{r-2M} + h + \xi\frac{d}{dr}\left( \Phi + \Lambda \right) +
\frac{2\xi}{r} + 2k + \frac{\partial \xi}{\partial r} \right]\\
\label{eq:ut_rot}
u^t  & = & e^{-\Phi}\left[ 1 - \xi\frac{d\Phi}{dr} - h +
\frac{1}{2}r^2\sin^2{\theta}{\bar{\omega}}^2e^{-2\Phi}\right],
\end{eqnarray}
where $m$, $h$, and $k$ are metric-perturbing functions which are
proportional to $\Omega^2$ and can be separated into spherical and
quadrupolar parts, $\bar{\omega}$ is the difference between the
star's rotation frequency and the
frame-dragging frequency, and $\xi$ is a Lagrangian displacement
that relates surfaces of constant $\rho$ in rotating and
non-rotating stars of the same $\rho_c$ \citep{hartle67}. The
spherical polar angle is $\theta$, and the mass enclosed inside
the radius $r$ is $M$. Replacing equations (\ref{eq:g_rot}) and
(\ref{eq:ut_rot}) in (\ref{eq:baryon_law}), keeping terms to order
$\Omega^2$, taking the angular integral, and integrating by parts
in $r$, we get\footnote{\citet{hartle67} and
\citet{hartlethorne68} give an expression for the total number of
baryons of the rotating star in which the last term of
equation~(\ref{eq:N_rot}) is not present, which is correct when
integrating up to the surface, but not at intermediate layers.}
\begin{eqnarray}
\label{eq:N_rot}
N(\rho_c, P, \Omega^2) & = & \int_0^{r(\rho_c, P)}4\pi y^2 e^{\Lambda(\rho_c,
y)} n(\rho_c, y) \left[ 1 + \frac{m_0}{y - 2M} + \frac{1}{3}y^2 \bar{\omega}^2
e^{-2\Phi}\right.\nonumber \\
 & & \left. - \frac{\xi_0(\rho_c,y)}{n(\rho_c,y)}\left( \frac{\partial
n}{\partial y}\right)_{\rho_c}\right]dy
 + 4\pi r^2 e^{\Lambda(\rho_c, r)} \xi_0(\rho_c,r) n(\rho_c, r).
\end{eqnarray}
The functions $\bar{\omega}$, $m_0$, $\xi_0$ are obtained by
solving first and second-order differential equations in the
coordinate $r$, as explained in \citet{hartle67}, and which need a
value for the rotation rate of the star $\Omega$ and a
non-rotating relativistic stellar structure as input. The zero
subscript on functions indicates their spherical part (the
quadrupolar part vanishes when performing the angular integral,
and the function $k$ has no spherical part).

%-----------------------------------------------------------
The remaining derivatives in equation (\ref{eq:dPdOmega2_final})
can be obtained from the non-rotating configuration. We first have
\begin{equation}
\label{eq:dNdP_def}
\left( \frac{\partial N}{\partial P}\right)_{\Omega^2, A} = \frac{4\pi r^2
e^{\Lambda} n}{dP/dr},
\end{equation}
where $dP/dr$ can be calculated analytically from the relativistic stellar
structure equations \citep{OV}.
The other derivative has to be obtained by differentiating equation
(\ref{eq:N_norot}):
\begin{eqnarray}
\label{eq:dNdn0_P}
\left( \frac{\partial N}{\partial \rho_c}\right)_{P, \Omega^2} & = & 4\pi r^2
e^{\Lambda(\rho_c,r)}n(\rho_c,r)\left( \frac{\partial r}{\partial
\rho_c}\right)_P\nonumber \\
 & & + \int_0^r 4\pi y^2\left[ e^{\Lambda(\rho_c,y)}\left( \frac{\partial
n}{\partial \rho_c}\right)_y + \frac{n(\rho_c,y)
e^{3\Lambda(\rho_c,y)}}{y}\left( \frac{\partial M}{\partial
\rho_c}\right)_y\right] dy.
\end{eqnarray}
For numerical calculations, it is convenient to write the
derivative in the first term of the right-hand side as
\begin{equation}
\label{eq:dPdn0_r}
\left( \frac{\partial r}{\partial \rho_c} \right)_P = -\frac{(\partial P /
\partial \rho_c)_r}{(\partial P / \partial r)_{\rho_c}},
\end{equation}
in order to calculate the three derivatives with constant $r$ in the same way.

The resulting spin-down compression rate, equation
(\ref{eq:dPdOmega2_final}), is plotted in
Figure~\ref{fig:dPdOmega2}, normalized by pressure and Kepler
frequency. The left panel displays results obtained for different
EOSs, with fixed central pressure, while the right panel shows
results for different stellar models and fixed EOS. As expected,
the expression is negative, since spin-down compression means
increase in $P$ with decrease of $\Omega^2$. Compression is
strongest at the center of the star. It can be shown that the
central value is given by
\begin{equation}
\left( \frac{\partial P}{\partial \Omega^2}\right)_{N,
A}\mathop{\longrightarrow}\limits_{r\to 0} -\frac{P_c \gamma_c (\rho_c +
P_c)}{\rho_c n_c} \frac{(\partial A/\partial \Omega^2)_{\rho_c}}{(\partial
A/\partial \rho_c)_{\Omega^2}},
\end{equation}
where $\gamma = d\log{P} / d\log{\rho}$ is the adiabatic index,
and the subscripts $c$ mean central values. Since $(\partial
A/\partial \rho_c)_{\Omega^2} = 0$ for the maximum mass
non-rotating configuration, the divergence of the result for $M\to
M_{max}$, as shown in the right panel of
Figure~\ref{fig:dPdOmega2}, is easy to understand. The graphs
confirm the order-of-magnitude expression
\begin{equation}
\left(\frac{\partial P}{\partial \Omega^2}\right)_{N,A}\sim
-\frac{P}{\Omega_K^2}
\end{equation}
which implies that the perturbation is small as long as $\Omega \ll \Omega_K$.

%-----------------------------------------------
With our expression for the spin-down compression rate at hand, we can write
the
change in the equilibrium number of particles of species $i$ with time as
\begin{equation}
\label{eq:Nieq}
\dot{N}_i^{eq} = 2\Omega \dot{\Omega}\int_{core} dN \frac{dY^{eq}_i}{dP}\left(
\frac{\partial P}{\partial \Omega^2} \right)_{N, A}
 \equiv 2\Omega \dot{\Omega} I_{\Omega, i},
\end{equation}
where the derivative $dY^{eq}_i/dP$ is calculated in beta equilibrium. It is
straightforward to check that
the $\dot{N}_i^{eq}$ satisfy baryon number and charge  conservation.

%---------------------------------------------------------------------------
\subsection{Effect of a Phase Transition}
\label{sec:transition}

The variables $N$ and $P$ are clearly continuous functions of $r$
in the non-rotating star, regardless of the presence of a phase
transition. In Appendix~\ref{sec:continuity}, we show explicitly
that the same is true for $(\partial P/\partial\Omega^2)_{N,A}$.
On the other hand, the equilibrium particle concentrations
may have a discontinuous jump $\Delta Y^{eq}_i$ at the transition,
which would contribute a Dirac delta function to $dY_i^{eq}/dP$.
Therefore, there would be a finite contribution to
$\dot{N}_i^{eq}$ from the transition,
\begin{equation}
\label{eq:Nieqtrans}
\dot{N}_i^{eq}|_{trans} = 2\Omega
\dot{\Omega}\Delta Y^{eq}_i \left[\frac{dN}{dP}\left(
\frac{\partial P}{\partial \Omega^2} \right)_{N,
A}\right]_{trans}.
\end{equation}

To order of magnitude, the fractional contribution of a phase
transition to the total value of $\dot{N}_i^{eq}$ (and to the
total heating of the star, if the reactions are of the same kind
everywhere) is the ratio of the jump $\Delta Y^{eq}_i$ to the
total variation in $Y^{eq}_i$ across the stellar core. In the
present work, as described in the following sections, phase
transitions are not important, as the only transition in one of
our EOSs is fairly weak\footnote{In our numerical integration, the
discontinuity is automatically taken care of by evaluating the
derivative $dY^{eq}_i/dP$ in each integration step as the ratio of
the (not necessarily small) increment in $Y^{eq}$ to the small
increment in $P$.}. In other cases, such as a star with a
core of deconfined quarks of fairly uniform relative abundances,
the phase transition may dominate the heating.

In order to maintain diffusive equilibrium, a finite number of
particles per unit time, $\dot{N}_i^{eq}|_{trans}$, must be
brought to (or, if negative, away from) the phase transition much
faster than the latter progresses due to the spin-down of the
star. Therefore, a phase transition is the place where our
assumption of diffusive equilibrium is most strongly put into
question. However, in the case of interest to us, the low
temperatures (implying long mean-free paths and therefore fast
diffusion) and the extremely slow evolution should still validate
this approximation.

\citet{glend92} has shown that, due to the presence of two
separate conserved charges (baryon number and electric charge),
phase transitions in neutron stars may not occur as a single
discontinuity, but instead through a mixed phase, in which both
phases coexist over a finite range in pressure (or stellar
radius). In this case, any discontinuities would be considerably
weakened, if not fully eliminated, and we do not expect any
special effects (such as discussed above) associated with the
phase transitions.

%%%%%%%%%%%%%%%%%%%%%%%%%%%%%%%%%%%%%%%%%%%%%%%%%%%%%%%%%%%%%%%%%%%%%%%%%%%%%
\section{INPUT FOR THE NON-SUPERFLUID CASE}

To implement the formalism of \S 2  numerically for the
non-superfluid case, we need as input an EOS, neutrino
emissivities, a specific heat, and an envelope model. We then have
to calculate the stellar structure and the corresponding integrals
over the star.

%-----------------------------------------
\subsection{Equations of State}
\label{sec:eos} Since our formalism requires a detailed knowledge
of the structure of the star, we need an EOS to determine the
spatial dependence of the thermodynamical quantities by solving
the relativistic stellar structure equations \citep{OV}. In
addition to the usual variables given in most published EOSs
(pressure, energy density, chemical composition, and adiabatic
index, all evaluated in chemical equilibrium), we need to know the
partial derivatives of equation~(\ref{eq:dni_dmui}) (which are
easy to compute if we know the energy density or energy per baryon
of interacting particles as a function of baryon number density
and proton fraction) and the baryon effective masses for computing
reaction rates and heat capacities. We also want to cover a broad
range of EOSs, in order to assess the dependence of the
quasi-equilibrium temperature on their properties. Given our
requirements, we have used three sets of EOSs:
\begin{enumerate}
\item the two most realistic EOSs from \citet*{apr98} (APR) for the core,
supplemented with that of \citet*{prl95} and
\citet{hanpichon94}, for the inner and outer crust, respectively;

\item five representative EOSs from \citet*{pal88} (BPAL) for the core, with
the same EOSs for the inner and
outer crust as for the previous set; and

\item a non-interacting Fermi gas (see, e.g., \citealt{shapiro}) for the whole
star.
\end{enumerate}
In Table~\ref{tab:eos_par} we list, for each EOS, the mass $M$,
central energy density $\rho_c$, coordinate radius $R$, and
effective radius as seen from infinity $R_\infty$ of the
maximum-mass non-rotating configuration.

Since in the APR set the thermodynamical quantities are obtained
from an analytical fit to tabulated many-body calculations
\citep{apr98}, care has to be taken when interpreting results
obtained for densities higher than the highest tabulated value,
since extrapolation can lead to significant errors. This is the
reason why we do not list the maximum-mass configuration for the
A18 + $\delta \upsilon$ EOS. Another feature of this set is that
the A18 + $\delta \upsilon$ + UIX* EOS becomes non-causal at
densities greater than $2\times 10^{15}\rm{g\,cm}^{-3}$, which is
the central density of a star of $2.14 M_\sun$, slightly below the
maximum mass formally allowed for this EOS, $M_{max}=2.19 M_\sun$.
However, this EOS is considered to be the state of the art among
those derived from non-relativistic potentials, since it
incorporates three-nucleon interactions and relativistic boost
corrections \citep{apr98}. For this reason, we will in what
follows take it as our reference EOS.

The EOSs of the BPAL set are labelled according to
\citet{prakash97},
who characterize them by the different magnitude of their nuclear
incompressibility and density dependence of their symmetry energy.
The values for the Fermi gas EOS listed in Table~\ref{tab:eos_par}
correspond to a central density slightly below the appearance of
$\Sigma^-$ hyperons.

The A18 + $\delta\upsilon$ + UIX* EOS has an additional feature: a
phase transition associated with the appearance of a neutral pion
condensate at a density $\sim 4\times 10^{14}$ g cm$^{-3}$.
In order to connect the normal phase with the pion-condensed one,
we assumed a Maxwell transition, which implies a discontinuity in
all thermodynamical quantities but the temperature, pressure and
neutron chemical potential, resulting in an energy-density jump of
6.6\%. \citet{apr98} also considered the possibility of a
mixed-phase transition (e.g., \citealt{glend92}), concluding that
(for a standard, $1.4M_\sun$ neutron star) the shell containing
the mixed phase is only $\sim 40$ m thick. They also estimated the
Debye screening length of electrons to be smaller than the
characteristic size of structures (by a factor 6-12), which would
favor the Maxwell transition as a better approximation to the real
situation \citep*{heiselberg93}.

Also listed in Table~\ref{tab:eos_par} are the Kepler periods $P_K
= 2\pi/\Omega_K$ corresponding to most of the EOSs, which are
necessary in order to determine the extent to which the
slow-rotation approximation holds for treating the spin-down
compression. The APR and PAL values were computed with the
empirical formula of \citet{lasota96}. This gives, for each EOS,
the maximum allowed rotation frequency in terms of the mass and
radius of the maximum-mass non-rotating configuration. The
accuracy for non-causal EOSs is somewhat lowered from its optimal
value of 1.5\% \citep{lasota96}, which applies for A18 +$\delta
\upsilon$ + UIX*. The reason why the entry corresponding to A18 +
$\delta \upsilon$ is empty is that its maximum mass configurations
occur at a central density which requires substantial
extrapolation, making the empirical formula an unreliable
estimator of $P_K$. For the Fermi gas EOS, we adopt the exact
value calculated for a pure neutron gas by \citet*{hanetal95},
since the empirical formula of \citet{lasota96} does not work for
this case. So far, the fastest known MSP is PSR B1937+21, with a
period of 1.56 ms \citep{bkh+82}, which is about three times
longer than Kepler periods for all realistic EOSs.

%--------------------------------------------------------------------
\subsection{Neutrino Emissivity}
\label{sec:reactions}

Since we are not taking superfluidity into account, direct and modified Urca
reactions
are the dominant neutrino emission mechanisms. We have therefore neglected
additional neutrino emission processes in the core and the crust of the star,
and ignored possible
modifications in Urca rates due to the neutral pion condensate of the APR set
(see, e.g., \citealt{khodel04}).

Away from beta equilibrium, neutrino emissivities and net reaction
rates per unit volume of Urca-type reactions in non-superfluid
matter are given by \citep{haensel92}
\begin{eqnarray}
\label{eq:Q_noneq}
Q_\alpha(n, T, \eta_{\alpha}) & = & Q^{eq}_{\alpha}(n, T)
F_*\left(\frac{\eta_{\alpha}}{kT}\right)\\
\label{eq:DGamma_noneq}
\Delta \Gamma_\alpha (n, T, \delta \eta_{\alpha}) & = &
\frac{1}{kT}Q^{eq}_\alpha(n, T) H_*\left(\frac{\eta_{\alpha}}{kT}\right),
\end{eqnarray}
respectively, where $Q_\alpha^{eq}$ is the neutrino emissivity in equilibrium
due to reaction $\alpha$,
$\eta_{\alpha}$ is the chemical imbalance affected by reaction $\alpha$, $T$ is
the local temperature,
$n$ is the baryon number density, $k$ is Boltzmann's constant, and
$F_*$ and $H_*$ are dimensionless control functions which depend on whether the
reaction is of direct of modified
Urca type. These functions are given by \citep{reisenegger95}
\begin{eqnarray}
\label{eq:F_D_def}
F_D(x) & = & 1 + \frac{1071x^2}{457\pi^2} + \frac{315x^4}{457\pi^4} +
\frac{21x^6}{457\pi^6}\\
\label{eq:H_D_def}
H_D(x) & = & \frac{714x}{457\pi^2} + \frac{420x^3}{457\pi^4} +
\frac{42x^5}{457\pi^6}\\
\label{eq:F_M_def}
F_M(x) & = & 1 + \frac{22020x^2}{11513\pi^2} + \frac{5670x^4}{11513\pi^4} +
\frac{420x^6}{11513\pi^6} + \frac{9x^8}{11513\pi^8}\\
\label{eq:H_M_def}
H_M(x) & = & \frac{14680x}{11513\pi^2} + \frac{7560x^3}{11513\pi^4} +
\frac{840x^5}{11513\pi^6} + \frac{24x^7}{11513\pi^6},
\end{eqnarray}
where the subscripts $D$ and $M$ mean direct and modified Urca, respectively. A
finite $\eta_\alpha$
of either sign enhances neutrino emission due to the even nature of the
functions $F_*$.
The amount of energy released by each reaction of type $\alpha$ is
$\eta_\alpha$ \citep{reisenegger95}, thus
the total energy dissipation rate per unit volume is
\begin{equation}
\label{eq:Q_H}
Q_H = \sum_\alpha \Delta \Gamma_\alpha \eta_\alpha,
\end{equation}
with the signs defined by
\begin{eqnarray}
\Delta \Gamma & = & \Gamma_{A\to B} - \Gamma_{B\to A}\\
\eta & = & \delta \mu(A) - \delta \mu(B),
\end{eqnarray}
where $\Gamma_{A\to B}$ is the rate per unit volume of the
reaction that transforms the particle set $A$ to the set $B$. This
sign convention ensures that reaction rates are enhanced in the
direction which restores equilibrium, since the functions $H_*$
are odd.

The equilibrium emissivities of both direct and modified Urca reactions can be
written as
\begin{equation}
\label{eq:Qeq}
Q_\alpha^{eq}(n, T) = S_\alpha(n)T^q,
\end{equation}
where $S_\alpha$ is a slowly varying function of $n$, $q=6$ for direct Urca,
and
$q=8$ for modified Urca (e.g., \citealt{yak01}). This decoupling
of the temperature from the spatial part introduces a great
simplification. To obtain the non-equilibrium neutrino
luminosities and net reaction rates, we first define
\begin{eqnarray}
\label{eq:xi_def}
\xi_\alpha & \equiv & \frac{\eta_\alpha}{kT} =
\frac{\eta_\alpha^\infty}{kT_\infty},\\
\label{eq:int_Qeq} \tilde{L}_\alpha & \equiv &
\frac{L^\infty_{\alpha,eq}}{T_\infty^q} = \int_{V_\alpha} 4\pi r^2
e^{\Lambda} S_\alpha(n) e^{\Phi(2 - q)} dr,
\end{eqnarray}
where we have used equations~(\ref{eq:T_int}), (\ref{eq:L_nu}) and
(\ref{eq:delta_mu_infty}), and where $L^\infty_{\alpha,eq}$ is the
equilibrium neutrino luminosity due to reaction~$\alpha$.
Combining equations~(\ref{eq:Q_noneq}), (\ref{eq:DGamma_noneq}),
(\ref{eq:xi_def}) and (\ref{eq:int_Qeq}), we obtain
\begin{eqnarray}
\label{eq:L_nu_j}
L^\infty_\alpha & = & \tilde{L}_\alpha F_*(\xi_\alpha) T_\infty^q,\\
\label{eq:int_DGamma}
\int \Delta \Gamma_\alpha e^{\Phi} dV & = & \frac{1}{k}\tilde{L}_\alpha
H_*(\xi_\alpha) T_\infty^{q-1}.
\end{eqnarray}
We denote modified Urca reactions with electrons and muons by
$\alpha$ = $Me$ and $\alpha = M\mu$, respectively, in each case
adding the contributions of the neutron and proton branches
\citep{yak01}. Direct Urca with electrons or muons is denoted by
$\alpha$ = $De$ and $D\mu$, respectively. From the APR set, the
only EOS which allows direct Urca is A18 + $\delta \upsilon$ +
UIX*, exceeding the threshold for direct Urca with electrons at
$\rho_{De} = 1.59\times 10^{15}$ g cm$^{-3}$, which is the central
density of a star of 2 $M_\sun$. Direct Urca with muons lies in
the non-causal regime. With the exception of BPAL31, all the EOSs
of the PAL set allow electron direct Urca. Muon direct Urca is
forbidden for both BPAL21 and BPAL31.

\begin{figure}
\plotone{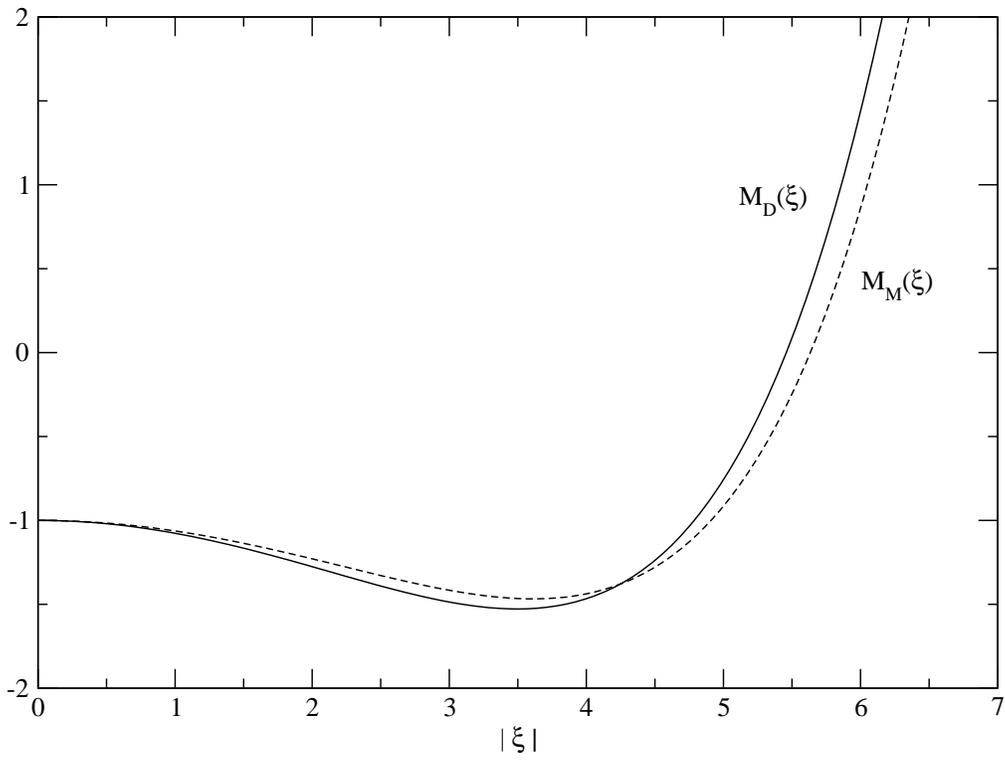} \caption{Functions $M_D$ and $M_M$, corresponding
to equation~(\ref{eq:M_j_def}) for direct and modified Urca
reactions, respectively. \label{fig:M_j}}
\end{figure}

Combining equations~(\ref{eq:L_H}), (\ref{eq:delta_mu_infty}),
(\ref{eq:DGamma_noneq}), (\ref{eq:Q_H}), (\ref{eq:xi_def}), and
(\ref{eq:int_DGamma}), we obtain the heating luminosity
corresponding to the reaction~$\alpha$
\begin{equation}
\label{eq:L_H_j} L^\infty_{H,\alpha} = \tilde{L}_\alpha \xi_\alpha
H_*(\xi_\alpha)  T_\infty^q.
\end{equation}
We can combine the heating and cooling contributions of each reaction into a
single expression, which will
prove to be useful for physical insight. If we define
\begin{equation}
\label{eq:M_j_def}
M_*(\xi_\alpha) = \xi_\alpha H_*(\xi_\alpha) - F_*(\xi_\alpha),
\end{equation}
we can write the difference of equations~(\ref{eq:L_nu_j}) and (\ref{eq:L_H_j})
as
\begin{equation}
\label{eq:net_cooling} L^\infty_{H,\alpha} - L^\infty_\alpha =
\tilde{L}_\alpha M_*(\xi_\alpha) T_\infty^q.
\end{equation}
Thus, the relative size of $\eta^\infty_\alpha$ and $kT_\infty$
will determine, through the functions $M_*$, whether there is net
cooling or heating due to the neutrino emitting reactions. Since
the functions $M_*$ are even, they do not depend on the sign of
the $\xi_\alpha$. We plot the functions $M_*(\xi)$ for the direct
and modified Urca case in Figure~\ref{fig:M_j}. With $\xi = 0$,
the constant term in $F_*$ gives the conventional cooling case. As
$|\xi|$ grows, the cooling is enhanced by the additional neutrino
emission due to the chemical imbalance,
reaching a maximum cooling at $|\xi| \sim 3.5$. For larger values
of $|\xi|$, the heating becomes important, completely balancing
the cooling for $|\xi| \sim 5.5$. For very large values of
$|\xi|$, the heating dominates, growing as $\sim \xi^6$ and $\sim
\xi^8$ in the direct and modified Urca case, respectively. In this
limit, a fixed fraction of the energy released is emitted as
neutrinos, 3/8 and 1/2 for modified and direct Urca, respectively.
The remainder stays behind, heating the star.

%-------------------------
\subsection{Envelope}
\label{sec:envelope}

Since we want to simulate the evolution of a MSP which has
finished accretion, we use the fully accreted envelope model of
\citet{potetal97} to calculate the effective surface temperature
\begin{equation}
\label{eq:envelope}
T_s^4 = 10^{24} g_{14} \left( 1.81 e^{-\Phi_b} T_{\infty,8} 
\right)^{2.42}\textrm{K}^4,
\end{equation}
where $g_{14}$ is the surface gravity in units of $10^{14}$ cm s$^{-2}$,
$T_{\infty,8}$ is the internal
temperature in units of $10^8$ K, and $\Phi_b \equiv \Phi(r_b)$.
The photon luminosity then follows from equation~(\ref{eq:L_gamma}). An
accreted envelope is more heat
transparent than a pure iron one for $T_s \gtrsim 10^5$ K, leading to faster
cooling, whereas for
$T_s\lesssim 10^5$ K the opposite happens \citep{potetal97}.

%--------------------------------------------------------------------
\subsection{Heat Capacity}

To calculate the integral in equation~(\ref{eq:capcal}), we use
the specific heat capacities at constant volume $c_{V,i}$ for
degenerate, non-superfluid fermions (see, e.g.,
\citealt{levyak94}). In analogy with the neutrino emissivity, we
can separate the temperature dependence from the spatial one. From
equation~(\ref{eq:capcal}), we can define
\begin{equation}
\label{eq:C_int}
\tilde{C} \equiv \frac{C}{T_\infty} = \frac{k^2}{3\hbar^3}\sum_i \int 4\pi r^2
e^\Lambda m_i^*(n) p_{Fi}(n) e^{-\Phi}dr,
\end{equation}
where $p_{Fi}$ and $m^*_i$ are the Fermi momentum and effective mass of
species~$i$, respectively. The latter
can be
obtained analytically for the APR and PAL EOSs (see, e.g., \citealt{page04}),
while for the noninteracting
Fermi gas we use $m^*_i = \mu_i/c^2$, where $\mu_i$ is the chemical potential
of species~$i$. The
sum is performed over all particle species and the integral is done over the
region where each particle
species exists. We take only free particles into account, neglecting the heat
capacity due to the lattice
of ions in the crust.

%-----------------------------------------------
\subsection{Crustal Processes}
\label{sec:crust}

In this work, we are neglecting processes in the crust which might
also modify the total number of neutrons, protons, or electrons.
We may ask here if this choice is a good approximation for
computing the evolution of the chemical imbalances, since the
short diffusion timescale limit means that the chemical
equilibrium everywhere in the star can be restored by reactions
occurring wherever these are fastest. \citet{iidasato97}
calculated the heating of a neutron star due to spin-down
compression in the crust. They found that neutron absorptions in
the inner crust are the dominant non-equilibrium process,
producing a very small heating rate $H(t)\approx 5\times 10^{-7}
\dot{E}$, where $\dot{E}$ is the spin-down power. Although they
did not consider neutron diffusion between different layers in the
star, and modelled the spin-down compression as a one-dimensional
process, we take their result to suggest a very small contribution
to the total heating. Since the inner crust accounts for about 2\%
of the total baryons in the star and the neutrino emissivities of
the core overwhelm crustal emission processes, we integrated
equations~(\ref{eq:Bij_def}) only over the core and neglected
processes in the crust, under the assumption that the error
introduced would be small. Our results confirm our assumption,
since for the more realistic EOSs we obtained heating rates due to
core processes much bigger than Iida \& Sato's results (see
equation \ref{eq:ratio_Edot}). We have to warn that neglecting the
crust may not be a good approximation if superfluidity is
included, since non-equilibrium processes in the crust can even
overtake suppressed Urca reactions in the core. It has also been
suggested by \citet{gusakov04} that direct Urca reactions could
occur at the crust-core interface. We have not explored this
possibility. However, we are aware that even a small amount of
direct Urca reactions in the inner crust would have a significant
effect on the evolution of chemical imbalances.

%--------------------------------------------
\subsection{Thermal Evolution Equations}
\label{sec:equations}

\begin{figure}
\plotone{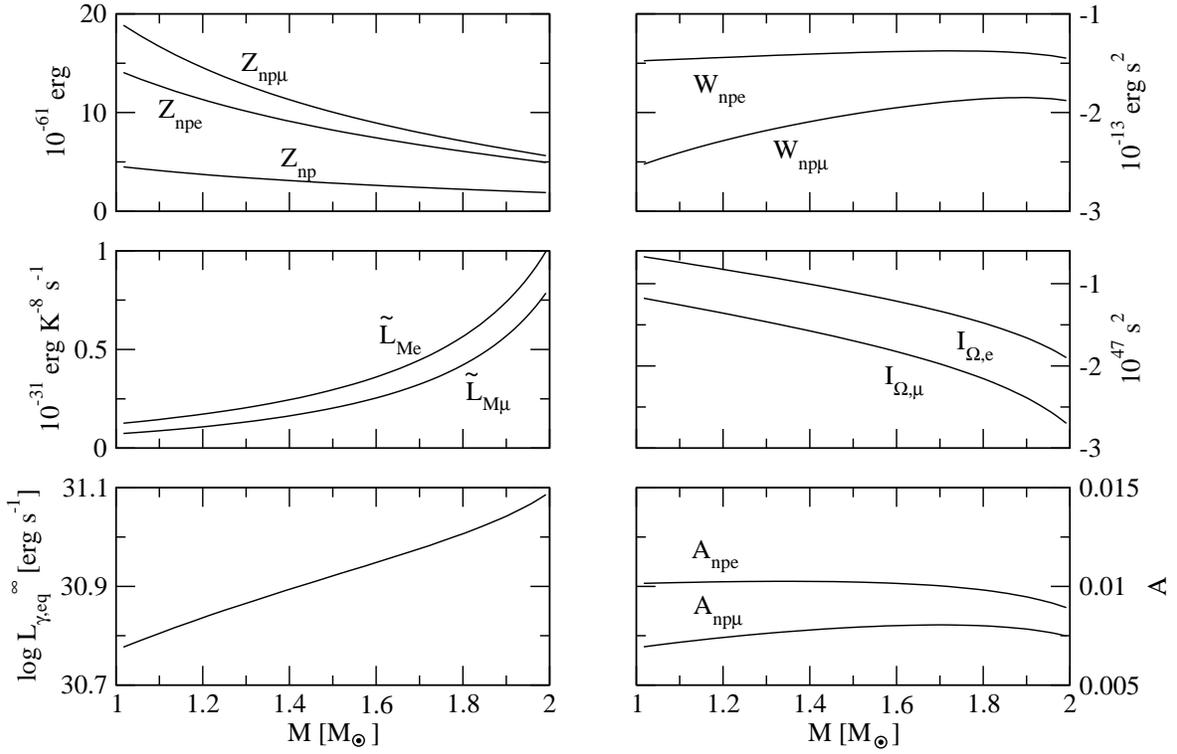} \caption{The four higher panels show results for
constants defined in section \S~\ref{sec:equations}, obtained from
different stellar models built with the A18 + $\delta \upsilon$ +
UIX* EOS. The two lower panels show the dependence on stellar
model of the quasi-equilibrium photon luminosity
(equation~\ref{eq:L_gamma_QE}) and the constant $A$, assuming
$P_{\rm{ms}} = \dot{P}_{-20} = 1$ and using the A18 + $\delta
\upsilon$ + UIX* EOS. For the latter, the two values are obtained
by taking $\ell=npe$ and $\ell = np\mu$ in
equation~(\ref{eq:A_def}), respectively.\label{fig:ctes}}
\end{figure}

Finally, we write the coupled equations for the time evolution of
the temperature and the chemical imbalances explicitly for the
non-superfluid case with $npe\mu$ composition, taking direct or
modified Urca reactions into account. For the internal
temperature, from equations~(\ref{eq:dot_T}), (\ref{eq:capcal}),
(\ref{eq:L_H}), (\ref{eq:L_nu}), (\ref{eq:L_gamma}),
(\ref{eq:net_cooling}), and (\ref{eq:C_int}), we get
\begin{eqnarray}
\label{eq:evolucion_Ti}
\dot{T}_\infty & = &\frac{1}{\tilde{C}}\left[
\left(M_D(\xi_{npe})\tilde{L}_{De} +
M_D(\xi_{np\mu})\tilde{L}_{D\mu}\right)T_\infty^5\right.\nonumber \\
 & & \left. + \left(M_M(\xi_{npe})\tilde{L}_{Me} +
M_M(\xi_{np\mu})\tilde{L}_{M\mu}\right)T_\infty^7 - L_\gamma^\infty\right].
\end{eqnarray}
For the chemical disequilibria, using equations~(\ref{eq:dNi_dmui}),
(\ref{eq:dNi_def}), (\ref{eq:dot_Ni}), (\ref{eq:dot_eta1}),
(\ref{eq:dot_eta2}), and (\ref{eq:Nieq}), we obtain
\begin{eqnarray}
\label{eq:evolucion_eta1}
\dot{\eta}^\infty_{npe} & = & -\frac{Z_{npe}}{k}\left(
\tilde{L}_{De}H_D(\xi_{npe})T_\infty^5 + \tilde{L}_{Me} H_M(\xi_{npe})
T_\infty^7\right) \nonumber \\
                    &   & - \frac{Z_{np}}{k}\left( \tilde{L}_{D\mu}
H_D(\xi_{np\mu}) T_\infty^5 + \tilde{L}_{M\mu} H_M(\xi_{np\mu})
T_\infty^7\right) + 2W_{npe}\Omega \dot{\Omega},\\
\label{eq:evolucion_eta2}
\dot{\eta}^\infty_{np\mu} & = & -\frac{Z_{np}}{k}\left(
\tilde{L}_{De}H_D(\xi_{npe})T_\infty^5 + \tilde{L}_{Me} H_M(\xi_{npe})
T_\infty^7\right) \nonumber \\
                    &   & - \frac{Z_{np\mu}}{k}\left( \tilde{L}_{D\mu}
H_D(\xi_{np\mu}) T_\infty^5 + \tilde{L}_{M\mu} H_M(\xi_{np\mu})
T_\infty^7\right) + 2W_{np\mu}\Omega \dot{\Omega},
\end{eqnarray}
where we have defined
\begin{eqnarray}
\label{eq:Z_np_def}
Z_{np} & = & \frac{B_{nn} + B_{np} + B_{pn} + B_{pp}}{B_{nn}B_{pp} -
B_{np}B_{pn}},\\
\label{eq:Z_npe_def}
Z_{npe} & = & \frac{1}{B_{ee}} + Z_{np},\\
\label{eq:Z_npm_def} Z_{np\mu} & = & \frac{1}{B_{\mu \mu}} +
Z_{np},
\end{eqnarray}
and
\begin{eqnarray}
\label{eq:W_npe_def}
W_{npe} & = & (Z_{npe}-Z_{np})I_{\Omega, e} + Z_{np}I_{\Omega, p},\\
\label{eq:W_npm_def}
W_{np\mu} & = & (Z_{np\mu}-Z_{np})I_{\Omega, \mu} + Z_{np}I_{\Omega, p}.
\end{eqnarray}
Typical values for the different constants are plotted in
Figure~\ref{fig:ctes}. Since $\Omega \dot{\Omega}<0$,
equations~(\ref{eq:evolucion_eta1}) and (\ref{eq:evolucion_eta2})
have a positive term, proportional to the spin-down power, which
arises from the change in the equilibrium concentrations of each
particle species with spin-down and makes the chemical imbalances
grow. The remaining terms, opposite in sign to $\eta_{npe}^\infty$
and $\eta_{np\mu}^\infty$, account for the effect of reactions
trying to restore beta equilibrium.
Equation~(\ref{eq:evolucion_Ti}) has the photon luminosity as a
negative term, and the net contribution of neutrino reactions,
equation~(\ref{eq:net_cooling}), which may be positive or negative
depending on the absolute value of $\xi_{npe}$ and $\xi_{np\mu}$.
For numerical calculations, the evolution of $\Omega \dot{\Omega}$
was computed assuming magnetic dipole braking with no field decay,
relating magnetic field, rotation period, and period derivative by
the conventional formula $B\simeq 3.2\times
10^{19}\sqrt{P\dot{P}}$ G, where $P$ is measured in seconds.

%%%%%%%%%%%%%%%%%%%%%%%%%%%%%%%%%%%%%%%%%%%%%%%%%%%%%%%%%%%%%%%%%%%%%%%%%%%%%%
\section{RESULTS AND DISCUSSION}

%------------------------------------------------------------
\subsection{Thermal Evolution}

\begin{figure}
\plottwo{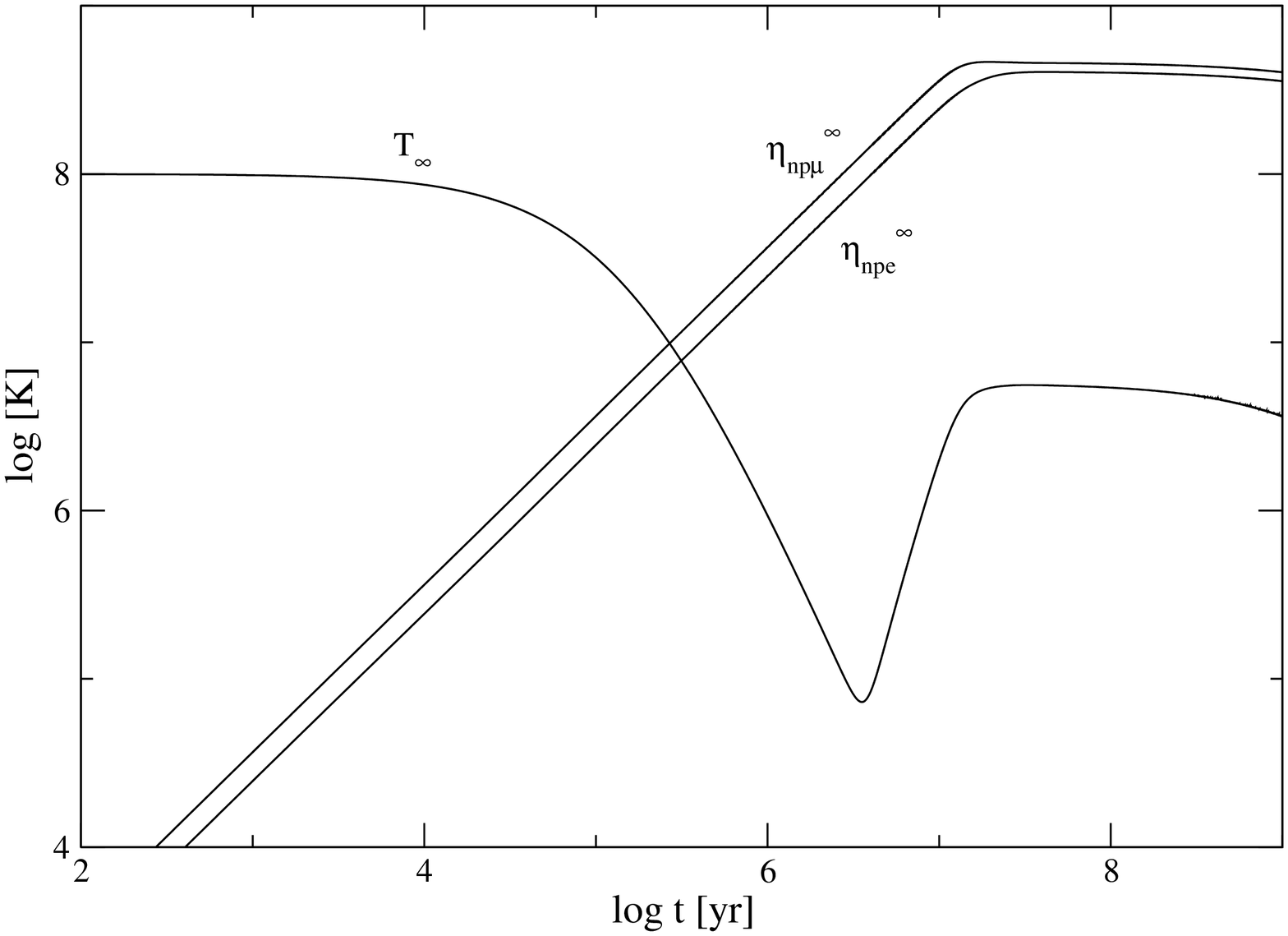}{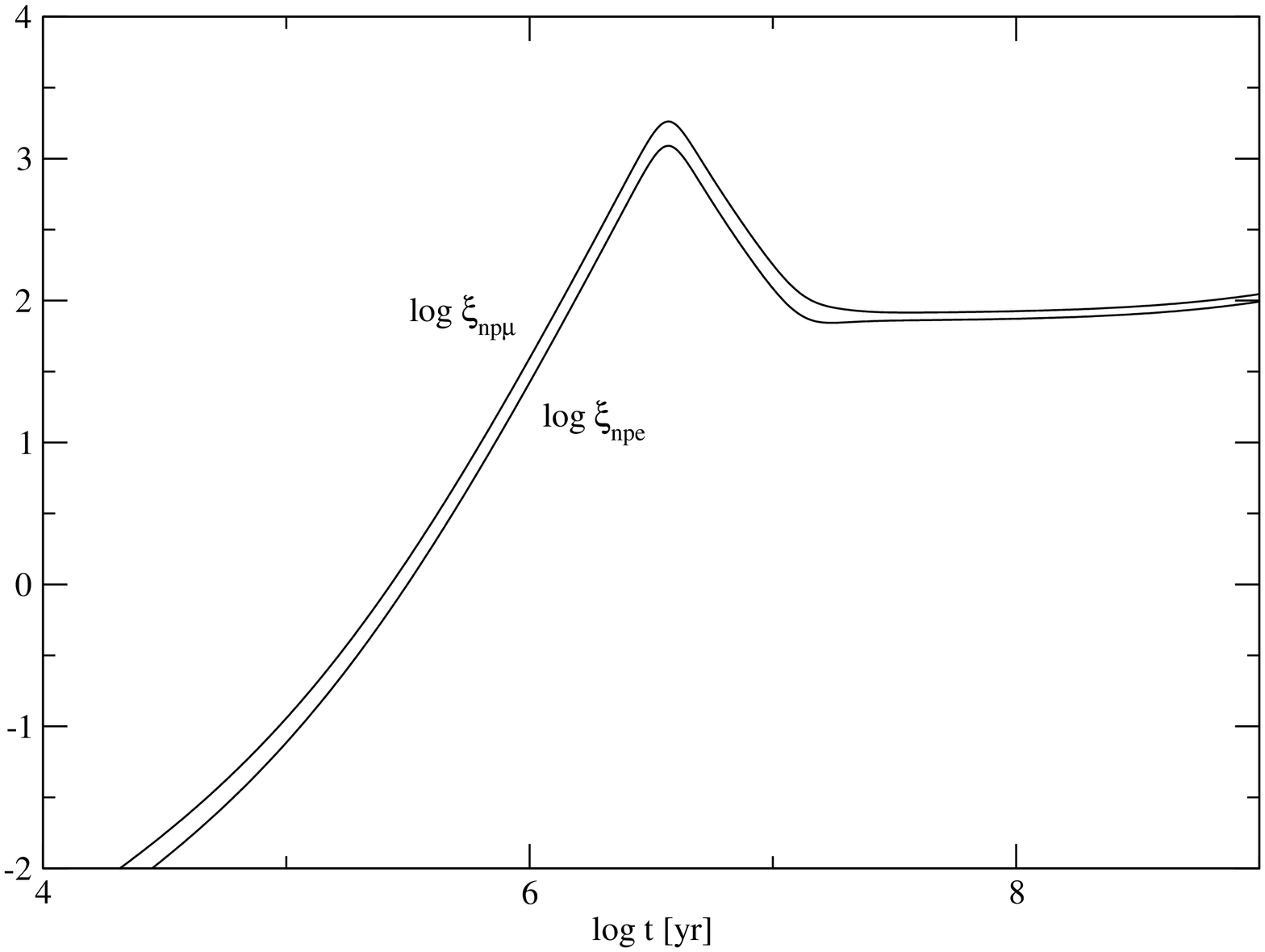} \caption{Evolution of the internal
temperature and chemical imbalances (left) and the ratios $\xi =
\eta^\infty/(kT_\infty)$ (right), for a 1.4$M_\sun$ star
calculated with the A18 + $\delta \upsilon$ + UIX* EOS, with
initial conditions $T_\infty = 10^8$ K and null chemical
imbalances at $t=0$, and spin-down parameters $B=10^8$ G and $P_0
= 1$ ms. \label{fig:evol_xi}}
\end{figure}

A typical solution of equations~(\ref{eq:evolucion_Ti}),
(\ref{eq:evolucion_eta1}), and (\ref{eq:evolucion_eta2}) is shown
in the left panel of Figure~\ref{fig:evol_xi}. First, the
temperature starts falling as in the conventional cooling case,
while the chemical imbalances grow due to spin-down compression.
At some point, the ratios $\xi_{npe}$ and $\xi_{np\mu}$ (right
panel of Figure~\ref{fig:evol_xi}) are big enough, so that the net
contributions of neutrino reactions,
equation~(\ref{eq:net_cooling}), are positive, and their sum
exceeds $L_\gamma^\infty$. The temperature then starts rising. The
chemical imbalances continue to grow, more slowly than the
temperature and hence reducing $\xi_{npe}$ and $\xi_{np\mu}$.
Finally, the star can arrive at a \emph{quasi-equilibrium} state,
where the rate at which spin-down modifies equilibrium
concentrations is the same as
the rate at which reactions drive the system towards the
equilibrium configuration, with heating and cooling balancing each
other \citep{reisenegger95}. The subsequent thermal evolution can
then be approximated by the simultaneous solution of the equations
$\dot{T}_\infty =0$ and $\dot{\eta}_{npe}^\infty =
\dot{\eta}_{np\mu}^\infty = 0$. We discuss this
\emph{quasi-equilibrium solution} in detail in \S4.2.

\begin{figure}
\plottwo{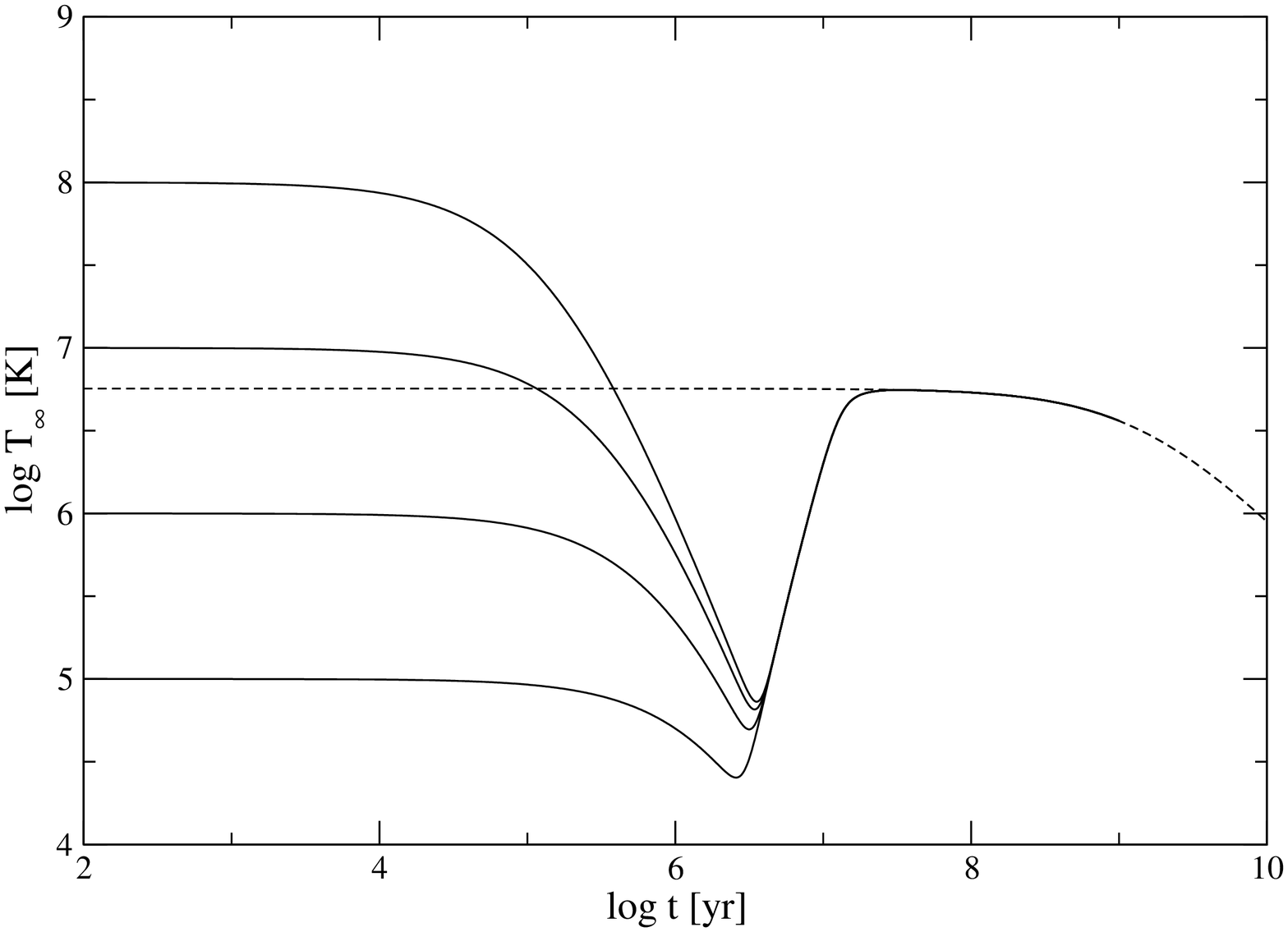}{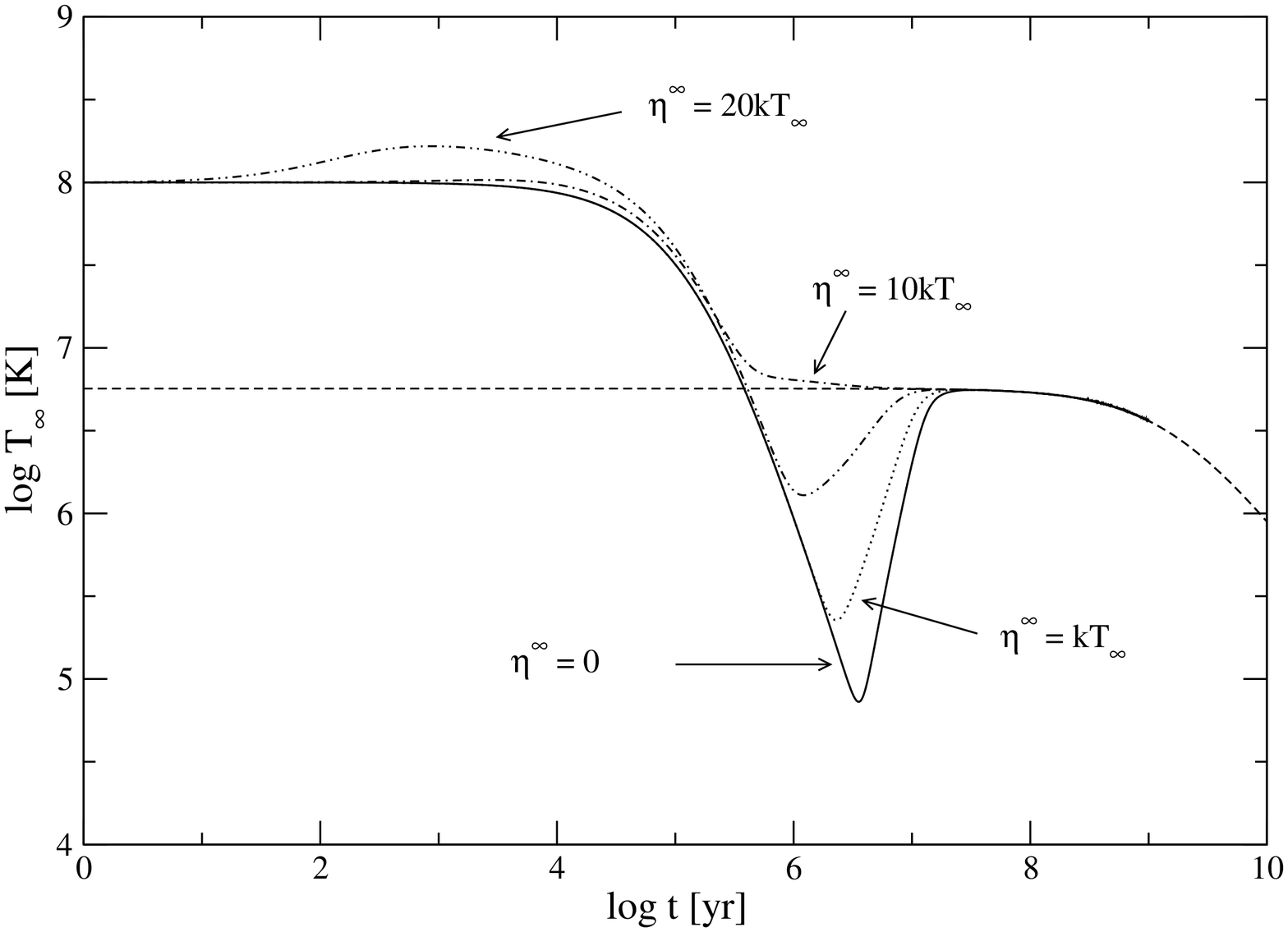} \caption{Evolution of the internal
temperature for different initial conditions on temperature (left)
and chemical imbalances (right). For both plots we set
$\eta_{npe}^\infty = \eta_{np\mu}^\infty \equiv \eta^\infty$ at
$t=0$. Fixed initial values are $\eta^\infty=0$ (left) and
$T_\infty = 10^8$ K (right). The short-dashed line is the
quasi-equilibrium solution, obtained by solving $\dot{T}_\infty
=0$ and $\dot{\eta}_{npe}^\infty = \dot{\eta}_{np\mu}^\infty =0$.
The stellar model and spin-down parameters are the same as in
Figure~\ref{fig:evol_xi}. \label{fig:evol_Tdmuvarios}}
\end{figure}

The pure modified Urca case ($\tilde{L}_{De} = \tilde{L}_{D\mu} =
0$) is the simplest to analyze. Figure~\ref{fig:evol_Tdmuvarios}
shows the thermal evolution of the same star as in
Figure~\ref{fig:evol_xi}, this time with different initial
conditions of internal temperature (left) and chemical imbalances
(right). The spin-down parameters were chosen so that the arrival
at the quasi-equilibrium state is clearly visible, the
short-dashed lines being the quasi-equilibrium solutions. We note
that, in both cases, the value of the temperature at the
quasi-equilibrium state and the time required to arrive do not
depend on initial conditions.

\begin{figure}
\plotone{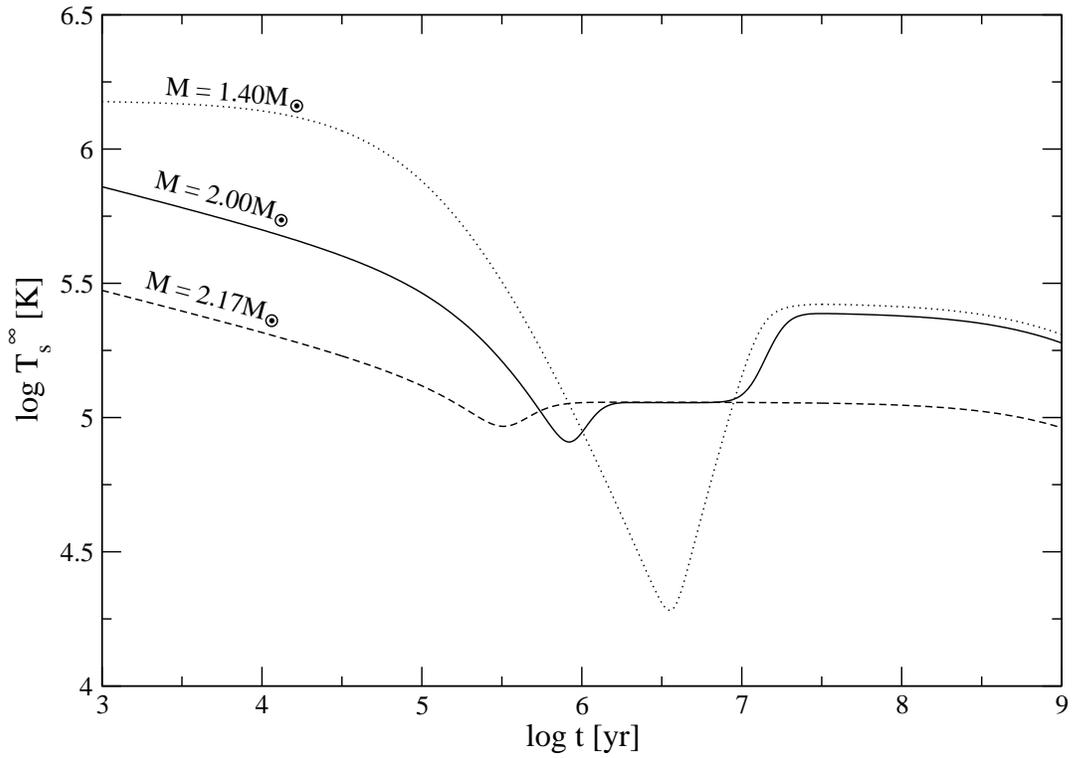} \caption{Evolution of surface temperature for
different stellar models, calculated using the A18 + $\delta
\upsilon$ + UIX* EOS, with fixed initial conditions. The mass of
each configuration is labelled on each curve. The 2$M_\sun$ star
is slightly above the threshold for direct Urca with electrons,
but below the threshold for muon direct Urca. The 2.17$M_\sun$
star is near the maximum-mass non-rotating configuration, and has
direct Urca with electrons and muons. The spin-down parameters are
the same as in Figure~\ref{fig:evol_xi}. \label{fig:evol_Urca}}
\end{figure}

If direct Urca reactions are open, solutions change noticeably. In
Figure~\ref{fig:evol_Urca} we show the thermal evolution of three
different stellar models built with the same EOS, with identical
initial conditions and spin-down parameters. The 1.4$M_\sun$ star
is the same as that of Figure~\ref{fig:evol_Tdmuvarios}, with only
modified Urca reactions in operation. The 2$M_\sun$ star has
electron direct Urca operating, with the corresponding muon
reaction forbidden. We see that the minimum temperature is higher
and occurs at a much earlier time, after which the system reaches
a ``metastable" quasi-equilibrium state, corresponding to partial
equilibration between $T_\infty$ and $\eta^{\infty}_{npe}$.
However, $\eta^{\infty}_{np\mu}$ continues to grow, leading the
system to a final quasi-equilibrium state, determined by muon
modified Urca. The 2.17$M_\sun$ star has direct Urca with
electrons and muons. This time the final quasi-equilibrium state
occurs even earlier than the ``metastable" quasi-equilibrium of
the 2$M_\sun$ star, with the temperature at about the same level.

%----------------------------------------------------
\subsection{The Quasi-Equilibrium State}

When only modified Urca reactions operate, it is possible to solve
analytically for the quasi-equilibrium values of the photon
luminosity $L_{\gamma,eq}^\infty$ and chemical imbalances
$\eta_{npe,eq}^\infty$ and $\eta_{np\mu,eq}^\infty$, as function
of stellar model and current value of $\Omega \dot{\Omega}$.
What makes these approximations possible is the high value of
$\xi_{npe}$ and $\xi_{np\mu}$ at
quasi-equilibrium. From the right panel of
Figure~\ref{fig:evol_xi}, we see that,
at quasi-equilibrium ($t\sim 10^{7.5}$ yr), $\xi_{npe}\sim
\xi_{np\mu}\sim 200$. This enables us to ignore all but the
greatest power in the functions $H_M$ and $M_M$ (equations
\ref{eq:F_M_def}, \ref{eq:H_M_def}, and \ref{eq:M_j_def}):
\begin{eqnarray}
\label{eq:C_H_def}
H_M(x) & \simeq & \frac{24}{11513\pi^8}x^7  \equiv  C_H x^7 \\
\label{eq:C_M_def}
M_M(x) & \simeq & \frac{15}{11513\pi^8}x^8  \equiv  C_M x^8.
\end{eqnarray}
The error in this approximation, for $x\sim 200$,
is less than one part in 100.

To solve for $L_{\gamma,eq}^\infty$, we set $\dot{T}_\infty =
\dot{\eta}^\infty_{npe} = \dot{\eta}^\infty_{np\mu} = 0$
and use
equations~(\ref{eq:C_H_def}) and (\ref{eq:C_M_def}) to rewrite
(\ref{eq:evolucion_Ti}), (\ref{eq:evolucion_eta1}), and
(\ref{eq:evolucion_eta2}) as
\begin{eqnarray}
\label{eq:dot_Ti_QE}
\tilde{L}_{Me} (\eta_{npe,eq}^\infty)^8 +  \tilde{L}_{M\mu}
(\eta_{np\mu,eq}^\infty)^8 & = & k^8 L_\gamma^\infty/C_M\\
\label{eq:dot_eta1_QE}
Z_{npe}\tilde{L}_{Me} (\eta_{npe,eq}^\infty)^7 + Z_{np}\tilde{L}_{M\mu}
(\eta_{np\mu,eq}^\infty)^7 & = & \frac{2k^8}{C_H} W_{npe} \Omega \dot{\Omega}\\
\label{eq:dot_eta2_QE}
Z_{np}\tilde{L}_{Me} (\eta_{npe,eq}^\infty)^7 + Z_{np\mu}\tilde{L}_{M\mu}
(\eta_{np\mu,eq}^\infty)^7 & = & \frac{2k^8}{C_H} W_{np\mu} \Omega \dot{\Omega}
\end{eqnarray}
Solving for $\eta_{npe,eq}^\infty$ and $\eta_{np\mu,eq}^\infty$ in
equations~(\ref{eq:dot_eta1_QE}) and (\ref{eq:dot_eta2_QE}), we get
\begin{eqnarray}
\label{eq:eta1_QE}
\eta_{npe,eq}^\infty  =  k\left(\frac{2 k I_{\Omega,
e}}{C_H\tilde{L}_{Me}}\right)^{1/7}(\Omega \dot{\Omega})^{1/7}\\
\label{eq:eta2_QE}
\eta_{np\mu,eq}^\infty  =  k\left(\frac{2 k I_{\Omega,
\mu}}{C_H\tilde{L}_{M\mu}}\right)^{1/7}(\Omega \dot{\Omega})^{1/7}
\end{eqnarray}
where we have used equations~(\ref{eq:W_npe_def}) and
(\ref{eq:W_npm_def}) and the fact that $I_{\Omega,p} =
I_{\Omega,e} + I_{\Omega,\mu}$. Both chemical imbalances at
quasi-equilibrium are positive, since $\Omega \dot{\Omega}$ and
$I_{\Omega, i}$ are negative. We finally replace equations
(\ref{eq:eta1_QE}) and (\ref{eq:eta2_QE}) in (\ref{eq:dot_Ti_QE}),
to obtain the photon luminosity:
\begin{equation}
\label{eq:L_gamma_QE}
L_{\gamma,eq}^\infty = C_M \left( \frac{2k}{C_H} \right)^{8/7}\left[ \left(
\frac{I^8_{\Omega, e}}{\tilde{L}_{Me}} \right)^{1/7} + \left(
\frac{I^8_{\Omega, \mu}}{\tilde{L}_{M\mu}} \right)^{1/7} \right]|\Omega
\dot{\Omega}|^{8/7}.
\end{equation}
An interesting feature of equation~(\ref{eq:L_gamma_QE}) is that
it does not depend on envelope model. The reason is that the limit
$\eta^{\infty}\gg kT_\infty$ implies that reactions are completely
determined by chemical imbalances, eliminating any dependence on
internal temperature. The energy released is divided in fixed
fractions between neutrino emission ($1 - [C_M/C_H] = 3/8$) and
heating ($C_M/C_H = 5/8$), the latter being radiated entirely as
photons. Note also that the ratio
$\eta_{npe,eq}^\infty/\eta_{np\mu,eq}^\infty$ depends only on EOS
and stellar model, and is of order unity, thus we will make
reference to either of them as $\eta^\infty_{\ell}$.

For the entire range of EOSs and stellar models, we can write
\begin{equation}
\label{eq:L_gamma_PLnum} L_{\gamma, eq}^\infty \simeq 10^{30-31}
\left(\frac{\dot{P}_{-20}}{P_{\rm{ms}}^3}\right)^{8/7}\textrm{erg
s}^{-1},
\end{equation}
where $\dot{P}_{-20}$ is the period derivative measured in units
of $10^{-20}$, and $P_{\rm{ms}}$ the period in ms. Values for
different stellar models built with the A18 + $\delta \upsilon$ +
UIX* EOS are shown in Figure~\ref{fig:ctes}. Using
equation~(\ref{eq:L_gamma}), we can rewrite
equation~(\ref{eq:L_gamma_PLnum}) in terms of the effective
temperature:
\begin{equation}
T_{s, eq}^\infty \simeq \left( 2 - 3\right)\times
10^5\left(\frac{\dot{P}_{-20}}{P_{\rm{ms}}^3}\right)^{2/7}\textrm{K}.
\end{equation}
This time, the uncertainty due to EOS and stellar model is smaller than 25\%.
We can also write the ratio of quasi-equilibrium photon luminosity (and thus
total heating rate) to the
spin-down power $\dot{E}$
\begin{equation}
\label{eq:ratio_Edot} \frac{L_{\gamma, eq}^\infty}{\dot{E}}\sim
(0.3-3)\times
10^{-5}\left(\frac{\dot{P}_{-20}}{P_{\rm{ms}}^3}\right)^{1/7}.
\end{equation}
It is also possible to calculate $\xi_{\ell,eq}$ for the quasi-equilibrium
state. Its spin-down power
dependence,
\begin{equation}
\xi_{\ell,eq} \propto |\Omega \dot{\Omega}|^{(\alpha - 8)/(7\alpha)},
\end{equation}
where $\alpha=2.42$ is the exponent of the envelope model of \citet{potetal97}
(see \S \ref{sec:envelope}),
shows that increasing $|\Omega \dot{\Omega}|$ reduces $\xi_\ell$, decreasing
the accuracy of the
analytical approximation, although very slowly.

%----------------------------------------------------
\subsection{Effect of Hyperons}

Although we did not include hyperons in our calculations, we may
ask how their presence could modify the thermal evolution. The
first particles to appear after muons are probably the
$\Sigma^{-}$ and $\Lambda^{0}$ hyperons, which require the
introduction of two additional chemical imbalances:
\begin{eqnarray}
\label{eq:eta_nnSp}
\eta_{nn\Sigma p} & = & 2\mu_n - \mu_\Sigma - \mu_p,\\
\label{eq:eta_nL}
\eta_{n\Lambda} & = & \mu_n - \mu_\Lambda.
\end{eqnarray}
Once hyperons are present, the following non-leptonic reactions
are open (e.g., \citealt{langer69}):
\begin{eqnarray}
\label{eq:nnSp}
n + n & \rightleftharpoons & \Sigma^- + p, \\
\label{eq:nnLn}
n + \Lambda^0 & \rightleftharpoons & \Sigma^- + p.
\end{eqnarray}
Reaction (\ref{eq:nnSp}) proceeds via weak interactions, since it
does not conserve strangeness, while (\ref{eq:nnLn}) proceeds via
strong interactions. Both reactions have timescales several orders
of magnitude shorter than beta processes \citep{langer69}.
Therefore, imbalances (\ref{eq:eta_nnSp}) and (\ref{eq:eta_nL})
remain small compared to $\eta_{npe}$ and $\eta_{np\mu}$, and
reactions (\ref{eq:nnSp}) and (\ref{eq:nnLn}) contribute
negligibly to the total heat generation. Moreover, since chemical
imbalances associated to direct or modified Urca reactions with
hyperons are linear combinations of $\eta_{nn\Sigma p}$,
$\eta_{n\Lambda}$, $\eta_{npe}$, and $\eta_{np\mu}$, the latter
two will determine the heat generation through both nucleon and
hyperon reactions.

In order to assess the importance of including the Urca processes
involving hyperons in addition to the nucleon processes, consider,
as an example, the $\Sigma^-$ direct Urca reactions
\begin{equation}
\Sigma^- \to n + e + \bar{\nu}_e,\qquad n + e\to \Sigma^- + \bar{\nu}_e,
\end{equation}
whose associated chemical imbalance is $\eta_{\Sigma n e} =
\eta_{npe} - \eta_{nn\Sigma p}\approx \eta_{npe}$. Their net
effect on equations~(\ref{eq:evolucion_eta1}) and
(\ref{eq:evolucion_eta2}) is to enhance the electron direct Urca
rate according to $\tilde{L}_{De}^{\prime} = \tilde{L}_{De} +
\tilde{L}_{\Sigma ne}$, reducing the chemical imbalance and thus
the stellar surface temperature. This correction is small, since
direct Urca reactions with hyperons are at least a factor 5 weaker
than their nucleon analogs \citep{prakash92}. For the pure
modified Urca case, it can be checked from
equation~(\ref{eq:L_gamma_QE}) that the correction to the surface
temperature introduced by adding several reactions involving
hyperons is a factor $[\tilde{L_n}/(\tilde{L}_n +
\tilde{L}_h)]^{1/28}$, where $\tilde{L}_n$ and $\tilde{L}_h$ are
the nucleon and hyperon Urca luminosities, respectively. Thus, the
corrections due to hyperons in a purely modified Urca or purely
direct Urca scenario are fairly negligible.

The only important effect of hyperons in the context of
rotochemical heating is that, in their presence, conditions for
direct Urca processes may become more easily satisfied, as is the
case with reactions involving $\Lambda^0$ \citep{prakash92}. The
steep increase of the proton concentration with the appearance of
$\Sigma^-$ hyperons (e.g., \citealt{glend97}) can also cause the
condition for nucleon direct Urca to be satisfied at lower
densities than if hyperons are excluded from the models.

%------------------------------------
\subsection{Conditions for Arrival at the Quasi-Equilibrium State}
\label{sec:conditions}

We may now ask which of the known MSPs are likely to be in the
quasi-equilibrium state. To answer this question, we need to know
how long it takes to arrive at the quasi-equilibrium state, and if
real pulsars are older than this time. Since the negative
``equilibration" terms due to reactions in
equations~(\ref{eq:evolucion_eta1}) and (\ref{eq:evolucion_eta2})
are important only for imbalances near their equilibrium values
(recall the limit $\xi_\ell \gg 1$), we can assume that the
initial evolution of $\eta^\infty_{\ell}$ is due only to $\Omega
\dot{\Omega}$:
\begin{equation}
\label{eq:etapto_aprox}
\dot{\eta}^\infty_\ell \approx 2W_\ell \Omega \dot{\Omega}.
\end{equation}
Integrating over time, we get
\begin{equation}
\label{eq:eta_ell_aprox}
\eta_\ell^\infty(t) \approx |W_\ell|\left[ \Omega^2_0 - \Omega^2(t) \right].
\end{equation}
For the true solution to have reached the quasi-equilibrium state,
this approximate solution must exceed
$\eta_{\ell,eq}^{\infty}(\Omega \dot{\Omega})$. Using
equations~(\ref{eq:eta1_QE}) or (\ref{eq:eta2_QE}), together with
equation~(\ref{eq:eta_ell_aprox}), we may express this condition
as an upper limit on the initial spin period
\begin{equation}
\label{eq:P0_qe}
P_0 < \frac{P}{\sqrt{1+A}}\equiv P_0^{qe},
\end{equation}
with
\begin{equation}
\label{eq:A_def} A = \frac{1}{|W_\ell|}\left(\frac{2k^8 I_{\Omega,
\ell}}{C_H \tilde{L}_\ell}\right)^{1/7} \frac{(\Omega
\dot{\Omega})^{1/7}}{\Omega^2}.
\end{equation}
(Here and in the rest of this subsection, we assume that only
modified Urca processes are active.)
Given the current spin parameters of a pulsar,
condition~(\ref{eq:P0_qe}) gives the highest initial period it can
have had, so that enough time has elapsed for it to have reached
the quasi-equilibrium state. For MSPs,
the constant $A$ is generally small. For a 1.4$M_\sun$ star built
with the A18 + $\delta \upsilon$ + UIX* EOS, we get
\begin{equation}
\label{eq:A_PPpto} A \approx 0.01 (\dot{P}_{-20}
P_{\rm{ms}}^{11})^{1/7}.
\end{equation}
Values for other stellar masses are shown in
Figure~\ref{fig:ctes}. From equation~(\ref{eq:etapto_aprox}), we
can estimate a characteristic timescale for equilibration as
\begin{equation}
\label{eq:tau_eq_def} \tau_{eq} \sim
\frac{\eta^\infty_{\ell,eq}}{2W_\ell \Omega \dot{\Omega}}
=1.6\times
10^7\left(P_{\rm{ms}}^3\over\dot{P}_{-20}\right)^{6/7}{\rm yr}.
\end{equation}

\begin{figure}
\plotone{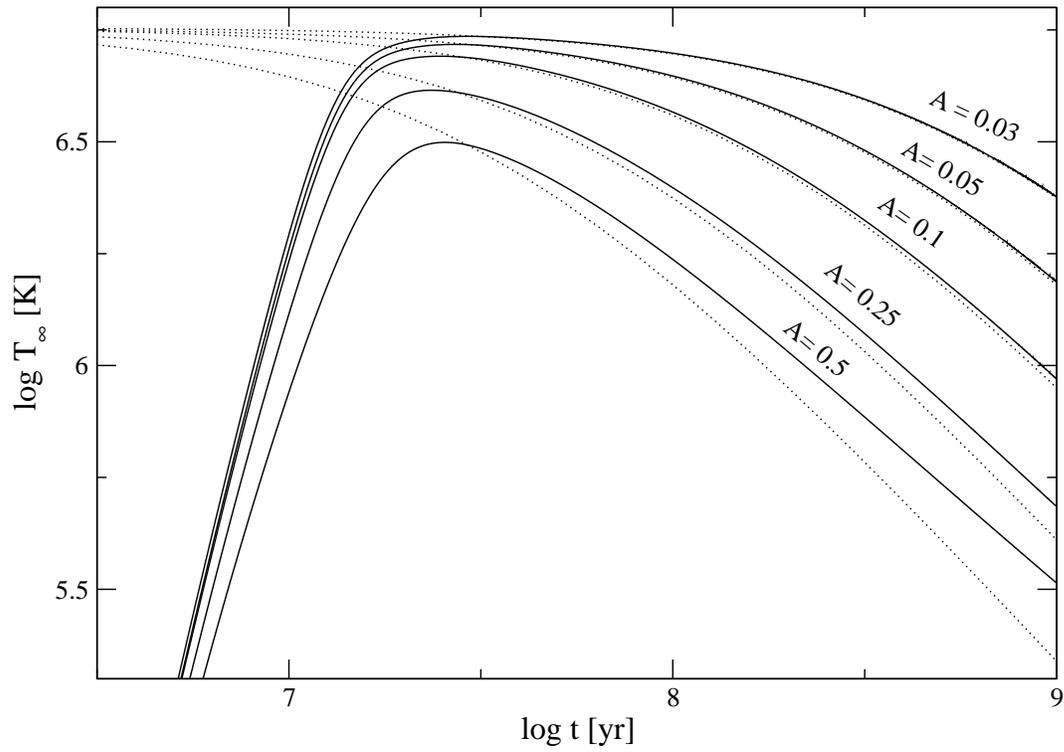} \caption{Exact and quasi-equilibrium solutions
for the internal temperature of the same star of
Figure~\ref{fig:evol_xi} (solid and dashed lines, respectively),
with fixed initial temperature, and different magnetic fields $B$,
choosing the initial period $P_0$ in each case so as to satisfy
$\tau_{eq}=1.5\times 10^7$ yr, and therefore implying different
initial values of $A$. \label{fig:evol_chiteqvarios}}
\end{figure}

\citet{reisenegger95} calculated the thermal evolution of neutron
stars with rotochemical heating for different values of the
magnetic field, using the magnetic dipole braking model with no
field decay. His conclusion was that a necessary condition for the
quasi-equilibrium solution to be a good approximation to the exact
one is that the spin-down timescale has to be longer than that for
the growth of chemical imbalances. To quantify this, we estimate
the spin-down timescale as the characteristic age,
\begin{equation}
\tau_{sd} = \frac{P}{2\dot{P}} = \frac{\Omega}{2\dot{\Omega}},
\end{equation}
and take the ratio of $\tau_{eq}$ to it, getting
\begin{equation}
\frac{\tau_{eq}}{\tau_{sd}} = A.
\end{equation}
Figure~\ref{fig:evol_chiteqvarios} shows the internal temperature
of our conventional star after the rise from the minimum
temperature, for different initial values of $A$, with fixed
initial $\tau_{eq}$. Dotted lines are the quasi-equilibrium
solutions at the instantaneous value of $\Omega\dot{\Omega}$. The
figure suggests that, for high values of $A$, exact and
quasi-equilibrium solutions depart from each other, whereas for
small $A$, solutions remain ``together''. In fact, all solutions
depart from each other, the departure being slower with time for
smaller $A$. For more general spin-down laws with arbitrary
braking index $n = \Omega \ddot{\Omega}/\dot{\Omega}^2$, it can be
shown analytically that this is true for $1<n<13$. The increase in
$\log{(T^{ex}/T^{qe})}$, where $T^{ex}$ is the exact solution and
$T^{qe}$ the quasi-equilibrium one, is roughly linear with time.
For $A=0.25$ at $t=0$, and assuming a standard braking index $n=3$,
$T^{eq}$ is $\sim 5\%$ lower than $T^{ex}$ at
$t=\tau_{eq}$, the difference growing 2\% every $10^8 $ yr.
Rewriting equation~(\ref{eq:A_PPpto}) in terms of the magnetic
field,
\begin{equation}
A \approx 0.01(P_{\rm{ms}}^5 B_8)^{2/7},
\end{equation}
where $B_8$ is the surface magnetic field in units of $10^8 $ G,
we can easily understand the results of \citet{reisenegger95},
since increasing $B$ with fixed $P_0$ increases the initial value
of $A$. Relating $A$ at the present time and at $t=0$ for an
arbitrary, but constant, braking index,
\begin{equation}
A(P,\dot{P}) = \left(\frac{P}{P_0}\right)^{(13-n)/7}A(P_0, \dot{P}_0),
\end{equation}
shows that the former is greater than the latter for all
reasonable values of $n$. Thus, a small current value of $A$ is a
good indicator that the star remains close to the
quasi-equilibrium state, as long as it satisfies $P_0 < P_0^{qe}$.

Since constraints on the initial periods $P_0^{wd}$ of a few MSPs
have been obtained from the cooling ages of their white dwarf
companions \citep{hansen98}, we can check which objects are likely
at the quasi-equilibrium state. In Table~\ref{tab:predictions},
further discussed in \S \ref{sec:otherMSPs}, we show the range in
$P_0^{wd}$ for some MSPs obtained by \citet{hansen98}. We made no
distinction between the $n=2$ and $n=3$ case,
and adopted the widest possible range in $P_0^{wd}$. In
Table~\ref{tab:predictions} we also show the value of $A$ and the
upper limit on initial spin period $P_0^{qe}$ required for the MSP
to be currently in the quasi-equilibrium state. Among
MSPs with constraints on initial periods, PSR J1012+5307 is the
only one not satisfying the conditions. The small values of $A$
imply that any initial periods substantially shorter than
the current ones will leave the MSPs in equilibrium by the present
time.

%------------------------------------
\subsection{Predictions for PSR J0437-4715 and PSR J0108-1431}

\begin{figure}
\plotone{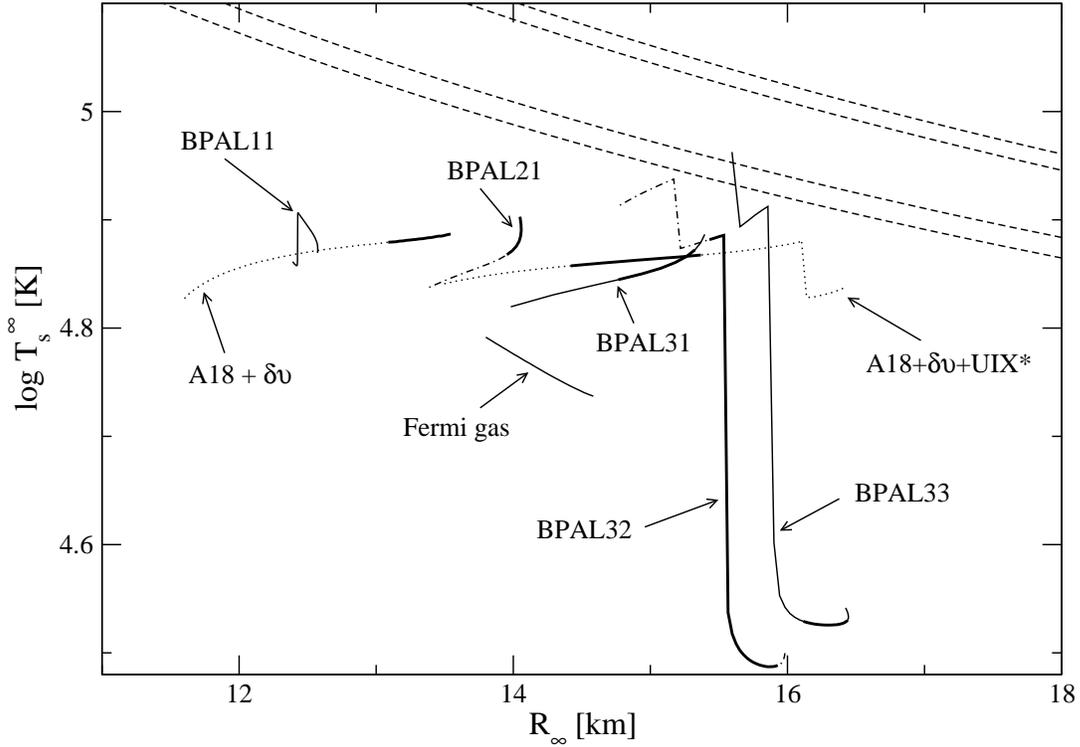} \caption{Quasi-equilibrium effective temperatures
$T_{s,eq}^\infty$ obtained with different EOSs and stellar models,
for the spin parameters of PSR J0437-4715. Dashed lines are the
68\% and 90\% confidence contours of the blackbody fit of
\citet{kargaltsev04} to probable thermal emission from this
pulsar. Bold lines indicate, for each EOS, the range corresponding
to the mass constraint of \citet{vbb01} for PSR J0437-4715,
$M_{PSR} = 1.58\pm 0.18M_\sun$. Abrupt reductions in temperature
with (increasing or decreasing) radius correspond to opening of
direct Urca reactions. \label{fig:Ts_Rinf}}
\end{figure}

Recently, \citet{kargaltsev04} measured ultraviolet emission from
PSR J0437-4715, the most probable explanation being that it
corresponds to thermal radiation. Since this object satisfies
$P_0^{wd}< P_0^{qe}$ with small $A$, we assume that it is already
at the quasi-equilibrium state. In Figure~\ref{fig:Ts_Rinf}, we
plot the blackbody fit of \citet{kargaltsev04} as dashed lines,
which correspond to 68\% and 90\% confidence contours. We overplot
our values of $T_{s,eq}^\infty$ for the spin parameters of PSR
J0437-4715 \citep{vbb01}, as function of $R_\infty$, for the EOSs
listed in Table~\ref{tab:eos_par}. The bold lines show the range
in $R_\infty$ corresponding to the mass constraint of
\citet{vbb01}, $M_{PSR} = 1.58\pm 0.18M_\sun$. The complete mass
range for each EOS goes from 1$M_\sun$ to 0.95$M_{max}$. The upper
limit was chosen to avoid the divergence in the spin-down
compression rate for masses near $M_{max}$ (see
Figure~\ref{fig:dPdOmega2}), which makes $T_{s,eq}^\infty$ also
diverge due to $I_{\Omega, i}$ (see equation~\ref{eq:L_gamma_QE}).

The best agreement with observations is reached if only modified Urca reactions
are present. When direct
Urca reactions
open, there are two abrupt drops in $T_{s,eq}^\infty$ with increasing stellar
mass, the first one
($\sim 10\%$) due to electron direct Urca, and the second ($\sim 50\%$) due to
muon direct Urca. This can be seen in
Figure~\ref{fig:Ts_Rinf} for the curves calculated with BPAL11 and A18 +
$\delta \upsilon$ + UIX*,
as examples of only electron direct Urca, and in the curves corresponding to
BPAL32 and BPAL33, which
have electron and muon direct Urca. Perfect agreement could be obtained by
allowing masses within
$\sim 1\%$ of the maximum-mass non-rotating configuration, for the EOSs with
only modified Urca.
However, we consider this scenario very unlikely.

Using the EOSs which have only modified Urca reactions within the allowed mass
for PSR J0437-4715, we can
constrain the predicted effective temperature to the narrow range
$T_{s,eq}^{\infty} = (6.9 - 7.9)\times 10^4$ K, about 20\% lower than the
blackbody fit of \citet{kargaltsev04}. There are three possible reasons why we
are not matching their
results:
\begin{enumerate}
\item  We are not taking superfluidity into account. This would reduce Urca
reaction rates,
lengthening the equilibration timescale and raising the quasi-equilibrium
temperature \citep{reisenegger97}.

\item We are neglecting other heating mechanisms (some of them directly related
to superfluidity), which
could further raise the temperature at any stage in the thermal
evolution. Nonetheless,
in MSPs, all proposed mechanisms are less important than
rotochemical heating \citep{schaab99}.

\item The thermal spectrum could deviate from a blackbody, in the same way as
the isolated neutron
star RX J1856-3754, which has a well-determined blackbody X-ray spectrum that
underpredicts the optical
flux \citep{walter}, indicating a more complex spectral shape of its thermal
emission.

\end{enumerate}

Kargaltsev and collaborators stress the fact that PSR J0437-4715,
despite its much larger spin-down age, has a higher surface
temperature than the upper limit for the younger, ``classical''
pulsar J0108-1431, $T_s^\infty<8.8\times 10^4$ K, inferred from
the optical non-detection by \citet{mignani03}. In this regard, we
note that the spin-down power ($\propto\Omega\dot\Omega$) of the
latter pulsar is 680 times lower than that of J0437, making
rotochemical heating substantially less important. Its
equilibration timescale, according to equation
(\ref{eq:tau_eq_def}), is $2\times 10^{11}$ yr, longer than the
age of the Universe and certainly much longer than the spin-down
age of the pulsar. Thus, it is not expected to be even close to
reaching its quasi-equilibrium state (although it is old enough to
have lost its initial heat content). Its actual temperature should
therefore be substantially smaller than the predicted
quasi-equilibrium surface temperature, which is already
$(680)^{2/7}\approx 6$ times lower than that of J0437, well below
the observational upper limit.

%------------------------------------
\subsection{Predictions for other millisecond pulsars}
\label{sec:otherMSPs}

In Table~\ref{tab:predictions}, we provide predictions of
$T_{s,eq}^\infty$ (assuming modified Urca reactions and no
superfluidity) for several MSPs. Since the high-energy part of
their expected thermal spectrum is highly absorbed by neutral
hydrogen in the interstellar medium, we chose to order them by
decreasing predicted quasi-equilibrium flux in the low-energy
(Rayleigh-Jeans) regime,
\begin{equation}
F_{RJ,eq} \propto \frac{T_{s,eq}^\infty}{d^2},
\end{equation}
where $d$ is the distance to the object. Each $F_{RJ,eq}$ entry in
Table~\ref{tab:predictions} is scaled by the value for PSR
J0437-4715. The first group of MSPs after PSR J0437-4715 are
nearby, single MSPs, for which $F_{RJ,eq}$ is expected to be
larger than for the binary MSPs, for which the initial period has
been estimated from their white dwarf companion (see \S
\ref{sec:conditions}). We consider the former
as the primary targets for observations analogous to those of
\citet{kargaltsev04}. However, even for PSR J0437-4715, the
nearest and brightest MSP known to date, the detection of thermal
emission is rather difficult, so the observation of other objects
of Table~\ref{tab:predictions} will be a real challenge. The
second group of MSPs are those with estimates for $P_0^{wd}$.
However, their low fluxes make the detection of their thermal
emission impossible for current instruments.

Optical detections are even more daunting than those in the
ultraviolet. The extrapolated Rayleigh-Jeans spectrum of PSR
J0437-4715 gives magnitudes $U=26.8$, $B=28.1$, and $V=28.5$,
which are completely overwhelmed by the white dwarf companion. The
most promising single MSPs have about three times lower expected
fluxes at Earth, i.~e., are another 1.2 magnitudes fainter, beyond
the reach of current telescopes.

%%%%%%%%%%%%%%%%%%%%%%%%%%%%%%%%%%%%%%%%%%%%%%%%%%%%%%%%%%%%%%%%%
\section{CONCLUSIONS}

In this work, we have made an extensive study of the effect of
rotochemical heating in non-superfluid millisecond pulsars. We
have set up a general formalism for the thermal evolution in the
framework of general relativity, which takes the stellar structure
fully into account and can be extended to include both
superfluidity and exotic particles. A key ingredient in this
formalism is the spin-down compression rate based on the
slow-rotation approximation of \citet{hartle67}.

The main consequence of rotochemical heating is the arrival at a
quasi-equilibrium state, in which the effective temperature of a
neutron star depends only on the current value of the spin-down
power. We argue that most of the known MSPs are very likely in the
quasi-equilibrium state. If only modified Urca reactions are
allowed, the quasi-equilibrium bolometric photon luminosity in the
non-superfluid case can be well approximated by
\begin{equation}
L_{\gamma,eq}^\infty \simeq
10^{30-31}\left(\frac{\dot{P}_{-20}}{P_{\rm{ms}}^3}\right)^{8/7}\textrm{erg
s}^{-1},
\end{equation}
independent of the neutron star envelope model.

The influence of EOS and stellar model on the quasi-equilibrium state is very
weak, the only significant
factor being the occurrence of direct Urca reactions. If they are open for both
muons and electrons,
quasi-equilibrium temperatures are low, as in the conventional cooling case. If
they are open only for
electrons, the system arrives at a ``metastable" quasi-equilibrium, after which
it proceeds as if only
modified Urca reactions with muons were present. The highest temperatures are
reached when all direct Urca
reactions are forbidden.

Even our highest predicted quasi-equilibrium effective temperatures are lower
than the blackbody fit
of \citet{kargaltsev04} to the UV emission of PSR J0437-4715 by about 20\%. The
inclusion of superfluidity
will likely raise our predicted temperatures, being the subject of future work.

\acknowledgments We thank G.~Pavlov and O.~Kargaltsev for letting
us know about their work in advance of publication and for kindly
providing the data for Figure~\ref{fig:Ts_Rinf}. The authors are
also grateful to M.~Taghizadeh, M.~van Kerkwijk, R.~Mignani,
D.~Page, M.~Catelan, M.~A.~D\'\i az, C.~Dib, and P.~Jofr\'e for
discussions that benefited the present paper, and an anonymous
referee for thoughtful comments that improved its final version.
This work made extensive use of NASA's Astrophysics Data System
Service, and received financial support from FONDECYT through
grant \# 1020840.

%%%%%%%%%%%%%%%%%%%%%%%%%%%%%%%%%%%%%%%%%%%%%%%%%%%%%%%%%%%%%%%%%%%
\appendix

%%%%%%%%%%%%%%%%%%%%%%%%%%%%%%%%%%%%%%%%%%%%%%%%%%%%%%%%%%%%%%%%%%%
\section{Coincidence of Surfaces of Constant Pressure and Energy Density in
General Relativity}
\label{sec:gradients}

%%%%%%%%%%%%%%%%%%%%%%%%%%%%%%%%%%%%%%%%%%%%%%%%%%%%%%%%%%%%%%%%%%%
%
% PLEASE DO NOT ALIGN SUBSCRIPTS AND SUPERSCRIPTS IN THIS APPENDIX
%
%%%%%%%%%%%%%%%%%%%%%%%%%%%%%%%%%%%%%%%%%%%%%%%%%%%%%%%%%%%%%%%%%%%

In this Appendix, we show that, in uniformly rotating,
relativistic stars in hydrostatic equilibrium, composed of a
perfect fluid, the surfaces of constant pressure and those of
constant energy density coincide, as in the Newtonian case, even
if the equation of state is not barotropic. In addition, the
gravitational redshift is constant on the same surfaces. This
result, together with the diffusive equilibrium condition, allows
us to describe the spin-down compression in terms of Lagrangian
changes of thermodynamic variables on isobaric surfaces enclosing
a fixed baryon number, since all thermodynamical quantities are
constant on each isobar. As a by-product, we also provide an
alternative derivation of the equation of hydrostatic equilibrium
for uniformly rotating, relativistic stars.

The assumption a perfect (i.e., non-dissipative) fluid, whose
stress-energy tensor depends only on the energy density $\rho$ and
(isotropic) pressure $P$, is of course not strictly true in the
astrophysical situation of interest in this paper, in which
entropy is generated by non-equilibrium weak interactions,
particle diffusion, and heat conduction, and lost (from the star)
through the emission of neutrinos and photons. However, the total
energy dissipated (and eventually lost) along the star's lifetime
is much smaller than that associated with the mass, random
motions, and interactions of its particles, as well as its
rotation, and the time scale of the release of this energy is many
orders of magnitude longer than the dynamical time and the
rotation period of the star. Therefore, the contribution of these
dissipative processes to the energy and momentum fluxes is
extremely small and can be neglected in the stress-energy tensor.
This is in line with the usual approach (also used in neutron star
cooling calculations and in the rest of this paper) of neglecting
thermal effects when calculating the structure of the star, which
is regarded as a fixed background on which thermal processes take
place.

The metric of a stationary, axially symmetric system can be
written in the form (see, e.g., \citealt{hartle67}):
\begin{equation}
ds^2 = g_{tt}dt^2 + g_{rr}dr^2 + g_{\theta \theta}d\theta^2 +
g_{\phi\phi}d\phi^2 + 2g_{t\phi}dtd\phi,
\end{equation}
where the metric coefficients are functions of $r$ and $\theta$
only. The components of the 4-velocity of any fluid element of the
uniformly rotating star are $u^r  =  u^\theta = 0$, and $u^\phi  =
\Omega u^t$, with
\begin{equation}
\label{eq:ut_def}
u^t = \left( -\left[ g_{tt} + 2\Omega g_{t\phi} + \Omega^2 g_{\phi
\phi}\right]\right)^{-1/2}.
\end{equation}
We note that $1/u^t$ is the gravitational redshift factor, which corrects
the energy of a freely moving particle, as it is
measured by observers co-moving with fluid elements on different
surfaces inside the star.\footnote{The locally measured energy is
$E=-u^\mu p_\mu=u^t(-p_t-\Omega p_\phi)$, where $p_\mu$ are the
covariant components of the particle's 4-momentum. Due to the
symmetries of the metric, the expression within parenthesis is
conserved along the particle's world line, and is its redshifted
energy.}

The energy-momentum conservation equation can then be written as
\begin{equation}
\label{eq:cons_enmom}
T^{\alpha \gamma}_{\phantom{\alpha \gamma};\gamma} = (\rho +
P)u^{\alpha}_{\phantom{\alpha};\gamma}u^{\gamma} + (\rho + P)u^{\alpha}
u^{\gamma}_{\phantom{\gamma};\gamma} + P_{,\gamma} g^{\alpha \gamma} = 0,
\end{equation}
where $T^{\alpha \beta}$ are the components of the stress-energy tensor of a
perfect fluid. Applying to
this equation a
projection operator orthogonal to the four-velocity,
$\wp_{\alpha \beta} = g_{\alpha \beta} + u_\alpha u_\beta$ \citep{schutz}, we
get
\begin{equation}
(g_{\alpha \beta} + u_\alpha u_\beta)(\rho + P)u^\gamma \Gamma^\alpha_{\delta
\gamma}u^\delta + P_{,\beta}=0.
\end{equation}
Since the only non-vanishing terms are those with $\gamma$ and
$\delta$ taking the values $\{t,\phi\}$, the relevant connection
coefficients are
\begin{equation}
\Gamma^\alpha_{\delta \gamma} = \cases{0, &if $\alpha=t, \phi$;\cr
-\frac{1}{2}g^{\alpha \mu}g_{\delta \gamma,\mu}, &if $\alpha = r, \theta$.\cr}
\end{equation}
Hence, equation~(\ref{eq:cons_enmom}) yields
\begin{equation}
P_{,\beta} - \frac{1}{2}(\rho + P)u^\gamma u^\delta g_{\delta \gamma, \beta} =
0.
\end{equation}
Using equation~(\ref{eq:ut_def}), we arrive at the equation of hydrostatic
equilibrium
\begin{equation}
\label{eq:hidroeq_rot}
P_{,\beta} - (\rho + P)(\ln{u^t})_{,\beta} = 0,
\end{equation}
which is in agreement with the result of \citet*{cook92},
for the special case of
rigid rotation. This shows that the
redshift factor $1/u^t$ is also constant on the surfaces of
constant $P$. Thus, we can write
\begin{equation}
\frac{dP}{d\ln{u^t}} = -(\rho + P),
\end{equation}
which implies that $\rho$ is also constant on the same surfaces.

%%%%%%%%%%%%%%%%%%%%%%%%%%%%%%%%%%%%%%%%%%%%%%%%%%%%%%%%%%
\section{Continuity of the Compression Rate across Phase Transitions}
\label{sec:continuity}

In this Appendix, we show that each term of the spin-down
compression rate, equation~(\ref{eq:dPdOmega2_final}), is
continuous across a phase transition,
even if the energy density and baryon number density change
discontinuously. We use units with $G = c = 1$.

The discontinuous quantities in the derivative $(dN/dP)_{\Omega^2,
A}$, equation~(\ref{eq:dNdP_def}), are $n$ and $dP/dr$. The Gibbs
free energy per unit volume for $npe\mu$ matter in beta
equilibrium at $T=0$ is
\begin{equation}
P + \rho = \mu_n n.
\end{equation}
Since, at a phase transition, the pressure $P$ and neutron chemical potential
$\mu_n$ are continuous,
the fractional size of the number density jump can be written as
\begin{equation}
\label{eq:dn_n}
\frac{\Delta n}{n} = \frac{\Delta \rho}{P + \rho}.
\end{equation}
where $\Delta n$ is the positive jump in number density. From the relativistic
stellar structure
equations, we have
\begin{equation}
\frac{dP}{dr} = -(\rho + P)\frac{M + 4\pi r^3 P}{r(r-2M)}.
\end{equation}
The fractional size of the corresponding jump across the phase
transition is
\begin{equation}
\label{eq:ddPdr_dPdr}
\frac{\Delta dP/dr}{dP/dr} = \frac{\Delta \rho}{\rho + P}.
\end{equation}
Thus, from equations~(\ref{eq:dn_n}) and (\ref{eq:ddPdr_dPdr}), we
see that the quantity $n/(dP/dr)$ is continuous across a phase
transition, and so is $(dN/dP)_{\Omega^2, A}$.

Regarding the derivative $(dN/d\Omega^2)_{P,\varepsilon_0}$,
equation~(\ref{eq:dNdOmega2_n0}), since the
jump in $n$ is integrated in
equation~(\ref{eq:N_norot}), $N(\rho_c, P, 0)$ is continuous. For $N(\rho_c, P,
\Omega^2)$,
the jump in $n$ due to the last term of equation~(\ref{eq:N_rot}) is cancelled
by the contribution from
$dn/dr$ in the integral. Using
\begin{equation}
\frac{dn}{dr} = -\Delta n \delta(r-r_t) + \textrm{continuous part},
\end{equation}
where $\delta$ is Dirac's delta function and $r_t = r(\rho_c, P_t)$, with $P_t$
the pressure at which the
phase transition occurs, we can write
\begin{equation}
\label{eq:disc_dNdOmega2}
-\int_0^{r} 4\pi y^2 e^{\Lambda(y)} \xi_0(y)\frac{dn}{dy}dy = 4\pi r_t^2
e^{\Lambda(r_t)} \xi_0(r_t) \Delta n \Theta(r-r_t) + \textrm{continuous part},
\end{equation}
where $\Theta$ is the step function. Equation~(\ref{eq:disc_dNdOmega2}) is
equal to minus the jump
(increasing $r$) of the last term in equation~(\ref{eq:N_rot}).

Finally, for the derivative $(dN/d\rho_c)_{P,\Omega^2}$, the contribution from
$(\partial n/\partial \rho_c)_r$ to the integral in equation~(\ref{eq:dNdn0_P})
cancels out the
discontinuity due to $n$ in the first term. Using
\begin{eqnarray}
\left(\frac{\partial n}{\partial \rho_c}\right)_r & = & \frac{r_i - r_t}{|r_i -
r_t|}\Delta n \delta(\rho_c - \rho_c^*) + \textrm{continuous part}\nonumber \\
 & = &\frac{r_i - r_t}{|r_i - r_t|}\Delta n \bigg|\left( \frac{\partial
r}{\partial \rho_c}\right)_{P}\bigg|_{r=r_t}\delta(r - r_t) +
\textrm{continuous part},
\end{eqnarray}
where $P(\rho_c^*, r) = P_t$, and $r_i(\rho_c)$ the radial coordinate at which
$(\partial n/\partial \rho_c)_r = 0$, we can write
\begin{equation}
\int_0^r 4\pi y^2 e^{\Lambda(y)}\left(\frac{\partial n}{\partial
\rho_c}\right)_{y} dy = 4\pi r_t^2 e^{\Lambda(r_t)} \frac{(r_i - r_t)}{|r_i -
r_t|}\Delta n \bigg|\left( \frac{\partial r}{\partial
\rho_c}\right)_{P}\bigg|_{r=r_t} \Theta(r-r_t) + \textrm{cont. part},
\end{equation}
which is the negative of the jump (increasing $r$) of the first term in
equation~(\ref{eq:dNdn0_P}),
since the sign of $(\partial r/\partial \varepsilon_0)_P$ is also $(r_i -
r_t)/|r_i - r_t|$.

\begin{figure}
\plottwo{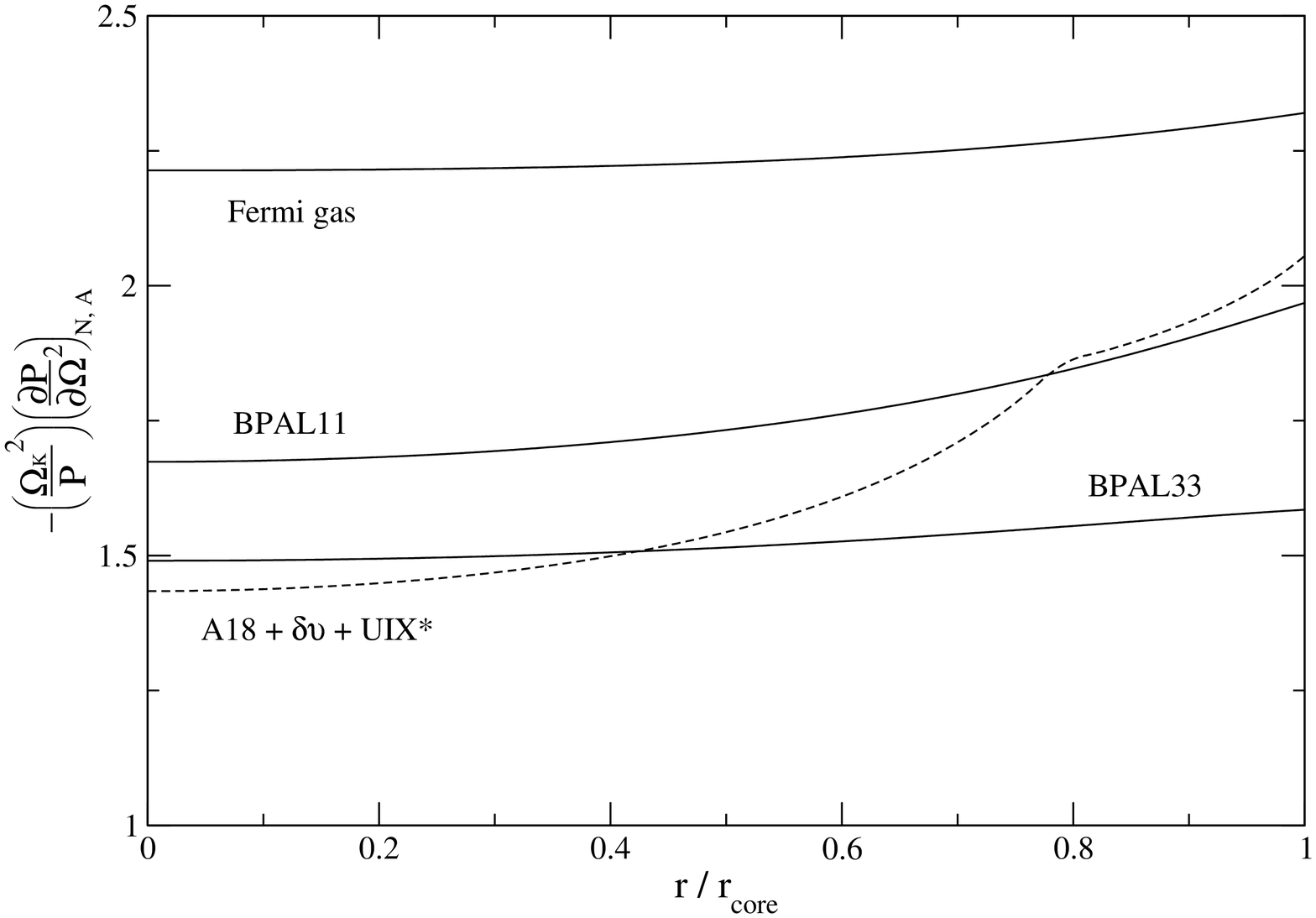}{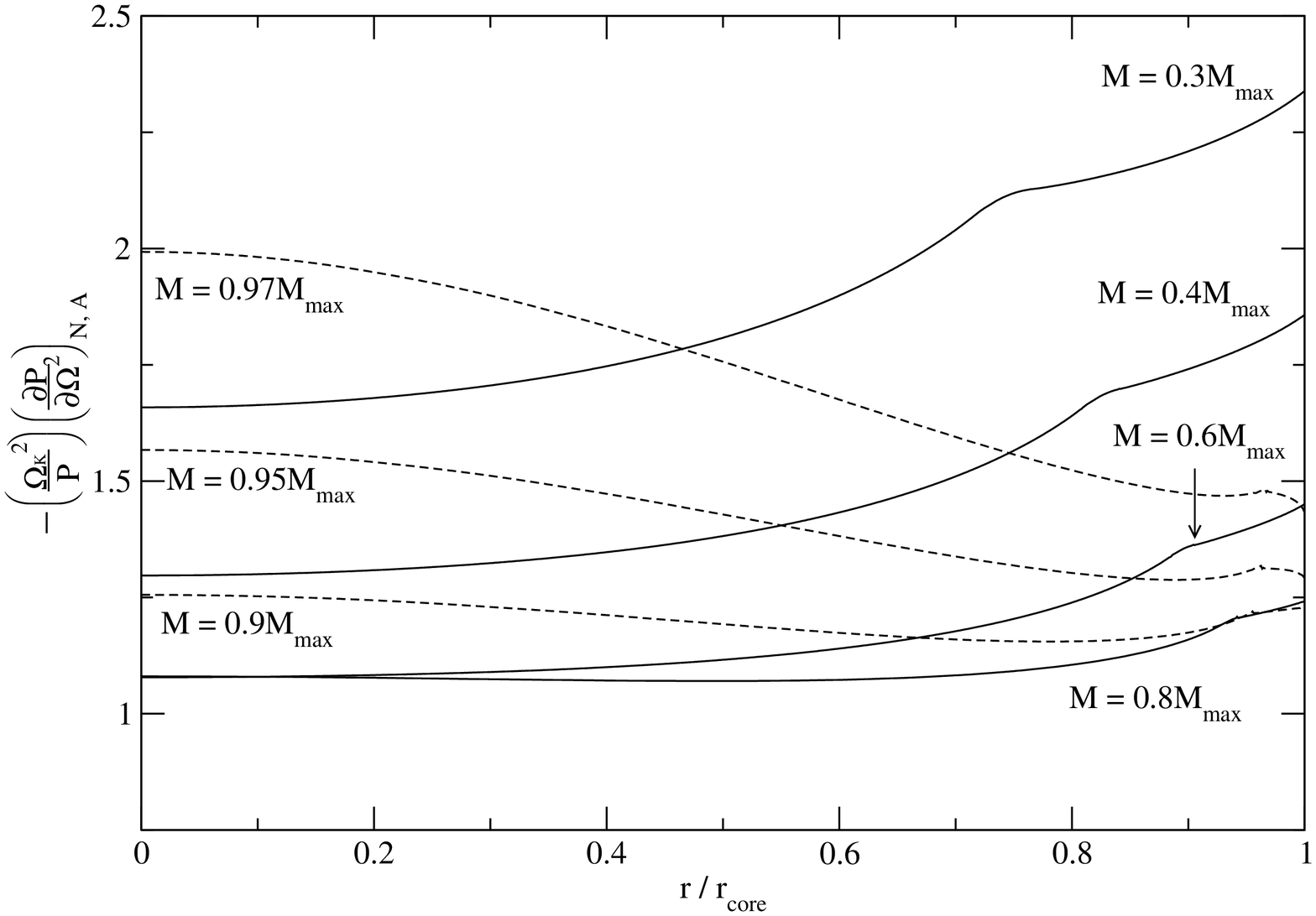} \caption{Left: Spin-down compression
rate for four different EOSs, normalized by pressure and Kepler
frequency. Stellar models have fixed central pressure $P_0 =
4.5\times 10^{34}$ dyn cm$^{-2}$. Right: Fractional compression
for different stellar models built with the A18 + $\delta
\upsilon$ + UIX* EOS. Stars are labelled by their mass in units of
the maximum-mass non-rotating configuration (2.19$M_\sun$).
\label{fig:dPdOmega2}}
\end{figure}

The right panel of Figure~\ref{fig:dPdOmega2} shows the spin-down
compression rate inside the core of several stellar models built
with the A18 + $\delta \upsilon$ + UIX* EOS, which has a phase
transition. The occurrence of the transition can be seen as a
(fairly harmless) discontinuity in the derivative of each curve.

%%%%%%%%%%%%%%%%%%%%%%%%%%%%%%%%%%%%%%%%%%%%%%%%%%%%%%%%%%%%%%%%%%%%%%

%%%%%%%%%%%%%%%%%%%%%%%%%%%%%%%%%%%%%%%%%%%%%%%%%%%%%%%%%%%%%%%%%%%

% dPdOmega2
%----------------

% Function M_j
%-----------------

% Constants
%---------------

\clearpage

% Evolution of eta and xi
%-----------------------

\clearpage

% Evolution with different initial conditions
%-----------------

% Evolution with direct Urca
%-----------------

% Evolution with different spin-down parameters
%-----------------

% Comparison with observations
%------------------

% EOS parameters
\begin{deluxetable}{lccccc}
\tablecaption{Maximum-mass non-rotating configuration and Kepler period for the
equations of state used
in this paper
\label{tab:eos_par}}
\tablewidth{0pt}
\tablehead{\colhead{EOS} & \colhead{$M_{max}$} & \colhead{$\rho_c$}            
     & \colhead{$R$}  & \colhead{$R_\infty$}  &
\colhead{$P_K$\tablenotemark{d}} \\
           \colhead{}    & \colhead{($M_\sun$)}                &
\colhead{($10^{15}$ g cm$^{-3}$)}   & \colhead{(km)} & \colhead{(km)}        &
\colhead{(ms)}
}
\startdata
A18 + $\delta \upsilon$           & 1.55\tablenotemark{a} & 1.86
&\phantom{1}9.81 & 13.42 &  -   \\
A18 + $\delta \upsilon$ + UIX$^*$ & 2.19\tablenotemark{b} & 2.78
&\phantom{1}9.97 & 16.79 & 0.51\\
BPAL 11                           & 1.42   & 4.45 &\phantom{1}8.42 & 11.86 &
0.54\\
BPAL 21                           & 1.70   & 3.46 &\phantom{1}9.33 & 13.69 &
0.56\\
BPAL 31                           & 1.91   & 2.86 &          10.10 & 15.18 &
0.59\\
BPAL 32                           & 1.95   & 2.66 &          10.56 & 15.64 &
0.63\\
BPAL 33                           & 1.97   & 2.53 &          10.92 & 15.94 &
0.66\\
Fermi gas                         & 0.62\tablenotemark{c}   & 1.10 &         
12.77 & 13.80 & 0.98
\enddata
\tablenotetext{a}{Corresponds to maximum value tabulated in
\citet{apr98}} \tablenotetext{b}{Stellar model lies in the
non-causal regime of this EOS} \tablenotetext{c}{Corresponds to
maximum mass before appearance of $\Sigma^-$ hyperons}
\tablenotetext{d}{Calculated with empirical formula (see text),
except the last value, which was adopted from \citet{hanetal95}}
\end{deluxetable}
%-----------------------

% Predictions for nearest MSPs
\begin{deluxetable}{lccccccccc}
\tablecaption{Predictions for MSPs likely to be observable or with estimates
for initial spin period
\label{tab:predictions}}
\tablewidth{0pt}
\tablehead{\colhead{Object} & \colhead{$P$}  &
\colhead{$\dot{P}$\tablenotemark{a}}    & \colhead{$d$\tablenotemark{b}}   &
\colhead{$T_{s,eq}^\infty$} & \colhead{$F_{RJ,eq}$}         &\colhead{$A$} &
\colhead{$P_0^{qe}$} &\colhead{$P_0^{wd}$} &\colhead{Refs.}\\
           \colhead{}       & \colhead{(ms)} & \colhead{($10^{-20}$)} &
\colhead{(kpc)} & \colhead{($10^5$ K)}        & \colhead{($F_{0437, eq})$}   
&\colhead{}  & \colhead{(ms)}       &\colhead{(ms)}       &\colhead{}
}
\startdata
J0437-4715  & 5.76 & 1.86 & 0.14 & 0.72 & 1    & 0.17 & 5.33 & 2.4--5.3 & 1,
2\\
\noalign{\bigskip}
J1024-0719  & 5.16 & $<$1.85 & $<$0.25 & $<$0.79 & $\sim$ 0.35\tablenotemark{c}
& $<$0.14 & $>$4.82 & \ldots   & 3, 4\\
J2124-3358  & 4.93 & 1.23 & 0.27 & 0.74 & 0.28 & 0.13 & 4.64 & \ldots   & 3,
4\\
J0030+0451  & 4.87 & $<$1.00 & 0.32 & $<$0.70 & $<$0.19 & $<$0.12 & $>$4.60 &
\ldots   & 4, 5\\
J1744-1134  & 4.07 & 0.71 & 0.36 & 0.74 & 0.15 & 0.09 & 3.90 & \ldots   & 6\\
B1257+12    & 6.22 & 4.26 & 0.45 & 0.86 & 0.12 & 0.22 & 5.63 & \ldots   & 3,
4\\
\noalign{\bigskip}
J0034-0534   & 1.88 & $<$0.67 & 0.53 & $<$1.41  & $<$0.14  & $<$0.03 & $>$1.85
& $<$1.4   & 2, 4, 7\\
J1012+5307  & 5.26 & 1.34 & 0.41 & 0.71  & 0.11  & 0.14 & 4.93 & $>$5.1   & 2,
4, 8\\
B1855+09    & 5.36 & 1.74 & 0.90 & 0.76  & 0.03  & 0.15 & 5.00 & $<$4.6   & 2,
3, 9\\
J1713+0747  & 4.57 & 0.81 & 1.10 & 0.70  & 0.02  & 0.11 & 4.35 & 2.2--3.1 & 2,
10\\
J1640+2224  & 3.16 & 0.16 & 1.15 & 0.60  & 0.01  & 0.05 & 3.08 &
$>$1.6   & 2, 4, 11\\
J2019+2425  & 3.93 & $<$0.70 & 1.49 & $<$0.76  & $<$0.01  &
$<$0.08 & $>$3.87 & 0.9--3.9 & 2, 4, 12
\enddata
\tablenotetext{a}{Intrinsic spin period derivative, based on the
latest available distance and proper motion. In cases where these
quantities are not known well enough for a reliable proper-motion
correction, the measured period derivative is given as an upper
limit. }

\tablenotetext{b}{Parallax distance when available, otherwise
dispersion-measure distance based on the NE2001 Galactic electron
density model \citep{ne2001}. In case of PSR J1024-0719, upper
limit based on measured proper motion and condition of positive
intrinsic period derivative. }

\tablenotetext{c}{Only rough reference value, since calculated as
ratio of two upper limits.}

\tablerefs{ (1) \citet{vbb01}; (2) \citet{hansen98}; (3)
\citet{Toscano99a}; (4) \citet{ne2001}; (5) \citet{Lommen}; (6)
\citet{Toscano99b}; (7) \citet{bhl+94}; (8) \citet{lcw+01}; (9)
\citet{ktr94}; (10) \citet{Splaver04}; (11) \citet{wdk+00}; (12)
\citet{nss01}.
}
\end{deluxetable}

%%%%%%%%%%%%%%%%%%%%%%%%%%%%%%%%%%%%%%%%%%%%%%%%%%%%%%%%%%%%%%%%%%%
\end{document}